\documentclass{aa}
\usepackage[usenames,dvipsnames]{xcolor}
\usepackage{natbib}
\usepackage{xspace}
\usepackage{amssymb}
\usepackage{amsmath}
\usepackage{graphicx}
\usepackage{hyperref}

\def\revised#1{{#1}}
\begin{document}
%\thesaurus{02.01.2,08.03.4,08.06.2,08.16.5,13.09.6}
\title{Cloudlet capture by Transitional Disk and FU Orionis stars}  \titlerunning{Cloudlet capture}
\authorrunning{Dullemond, K\"uffmeier, Goicovic, Fukagawa, Oehl, Kramer} 
\author{C.P.~Dullemond$^{1}$, M.~K\"uffmeier$^{1}$, F.~Goicovic$^{1}$, M.~Fukagawa$^{2}$, V.~Oehl$^{1}$, M.~Kramer$^{1}$}
\institute{
  (1) Zentrum f\"ur Astronomie, Heidelberg University, Albert Ueberle Str.~2, 69120 Heidelberg, Germany\\
  (2) Division of Particle and Astrophysical Science, Graduate School of Science, Nagoya University, Furo-cho, Chikusa-ku, Nagoya, Aichi 464-8602, Japan
} \date{\today}

\abstract{After its formation, a young star spends some time traversing the
  molecular cloud complex in which it was born. It is therefore not unlikely
  that, well after the initial cloud collapse event which produced the star, it
  will encounter one or more low mass cloud fragments, \revised{which we call
  ``cloudlets'' to distinguish them from full-fledged molecular clouds.} Some of
  this cloudlet material may accrete onto the star+disk system, while other
  material may fly by in a hyperbolic orbit. In contrast to the original cloud
  collapse event, this process will be a ``cloudlet flyby'' and/or ``cloudlet
  capture'' event: A Bondi-Hoyle-Lyttleton type accretion event, driven by the
  relative velocity between the star and the cloudlet. As we will show in this
  paper, if the cloudlet is small enough and has an impact parameter similar or
  less than $GM_{*}/v^2_\infty$ (with $v_\infty$ being the approach velocity),
  such a flyby and/or capture event would lead to arc-shaped or tail-shaped
  reflection nebulosity near the star. Those shapes of reflection nebulosity 
  can be seen around several transitional disks and FU Orionis stars. Although
  the masses in the those arcs appears to be much less than the disk masses in
  these sources, we speculate that higher-mass cloudlet capture events may also
  happen occasionally. If so, they may lead to the tilting of the outer disk,
  because the newly infalling matter will have an angular momentum orientation
  entirely unrelated to that of the disk.  This may be one possible explanation
  for the highly warped/tilted inner/outer disk geometries found in several
  transitional disks. We also speculate that such events, if massive
    enough, may lead to FU Orionis outbursts.  }

\maketitle


\section{Introduction}

It is well known that stars rarely form in isolation. Star formation happens in
Giant Molecular Cloud complexes (GMCs), producing tens to many thousands of stars
before the molecular cloud dissipates. The process is most likely primarily
regulated by turbulence \citep{2004RvMP...76..125M}. The complex and chaotic
process by which such turbulent cloud complexes produce stars can be modeled in
quite some detail with 3-D (magneto-)(radiation-)hydrodynamic simulations
\citep[e.g.][]{2016A&A...591A..30L}. Zoom-in simulations can follow this process
from the lagest scales all the way down to the detailed dynamics and evolution
of the protostellar/protoplanetary disk \citep[e.g.][]{2017ApJ...846....7K}.
From such simulations it is known that the formation of a star in such an
environment usually occurs in fits and starts rather than in a smooth single
collapse, and that during these accretion events the angular momentum of the
infalling material can change \citep{2004A&A...423....1J}. In spite of these
complexities, as a rule of thumb the formation of a star is usually regarded as
a three-stage process: At first, a molecular cloud core becomes gravitationally
unstable and starts to collapse, forming a protostar in the center (the class 0
stage). This protostar will initially be still surrounded by the infalling
envelope material, which feeds the star and its protostellar disk (the class I
stage). Once the collapse is over, the star and its protoplanetary disk are
revealed (the class II stage). Such a simplification can be a powerful
tool to understand the overall processes through which stars and their disks
form and what properties they have \citep[see e.g.][]{2005A&A...442..703H,
  2006ApJ...645L..69D}. But it obviously also has its limitations.

For the computation of the Initial Mass Function from 3-D star formation
simulations it suffices to focus on the main infall phase, where most of the
mass is accreted onto the star+disk system. Although the simple three-stage
scenario mentioned above is too simplified, it is in general the case that most
of the mass accretion is over by about one free-fall time scale. However, if we
are interested in the proto{\em planetary} disk and its evolution, even
relatively low-mass ($\delta M\ll M_{\odot}$) accretion events occurring well after the main core collapse
phase may become relevant, because these disks are thought to have masses of
only $10^{-3}$ to at most $\sim 10^{-1}$ $M_{\odot}$. It is very natural to expect
such late-time low-mass accretion events to take place, because after its
formation, the ``finished'' star+disk system continues to travel through the
remainder of the GMC. It may thus continue to accrete
gas and dust from cloud fragments \revised{(henceforth called ``cloudlets'')} at a low rate through the
Bondi-Hoyle-Lyttleton process (hereafter BHL process).
\citet{2014A&A...566L...3S} argue that this might explain ``late accretors'':
stars that are relatively old (up to 30 Myr) but still show substantial
accretion rates. \citet{2005ApJ...622L..61P} and \citet{2008AJ....135.2380T}
propose that the BHL process may be (in part) responsible for the observed $\dot
M\propto M_{*}^{1.8}$ relation of pre-main-sequence stas \citep{2003ApJ...592..266M,
  2004A&A...424..603N}. \citet{2010A&A...520A..17K} generalize this idea to a
lower stellar velocity dispersion of $\sim$0.2 km/s, which they expect, instead
of the canonical $\sim$1 km/s, and show that under those conditions even relatively
high accretion rates can be obtained.

Since the Interstellar Medium in general and GMCs in
particular have overall a fractal, clumpy structure, such late-time BHL
accretion is presumably highly intermittent. Instead of encountering smooth
large scale cloud structures, the star+disk system will likely encounter
cloudlets of various densities and sizes. Cloudlets of sizes down to few
hundreds (or even a few tens) of AU, and masses much less than a solar mass have
been observed \citep[][]{1995ApJ...453..293L, 1997ApJ...481..193H,
  2004Ap&SS.292...89F, 2012ApJ...754...95T}. These cloudlets are clearly stable
against gravitational collapse and will not form stars themselves. Only through
BHL accretion they may contribute to the mass of stars and their disks, if
they pass too close by one of the young stars.

If the size of a cloudlet is of the same order as, or smaller than, the
Hoyle-Lyttleton radius $R_{\mathrm{HL}}\equiv 2GM_{*}/v_{\infty}^2$ (where
$v_{\infty}$ is the relative velocity of the cloudlet with respect to the star),
it becomes important to consider the precise impact parameter $b$ at which the
cloudlet approaches the star. A close enough approach of such a cloudlet to a star
could lead to some of the matter remaining gravitationally bound to the star. We
will henceforth call this process a ``cloudlet capture event''. Statistically a
perfect head-on ``collision'' ($b=0$) is unlikely, while a very large impact
parameter ($b\gg R_{\mathrm{HL}}$) will not lead to any matter getting
captured. Therefore cloudlet capture events will most commonly occur with impact
parameters similar to the Hoyle-Lyttleton radius.

As a result, if the cloudlet (or part of it) is indeed captured by the star, it
carries along a very high specific angular momentum with respect to the star.
The captured material will thus not directly fall onto the star itself but will
orbit around it. If the star does not have (or no longer has) a disk, then this
leads to the formation of a new disk (or secondary disk). If a disk pre-exists,
then the infalling matter will, in all likelihood, have a very different angular
momentum axis than the disk. Depending on the mass ratio of the infalling
cloudlet and the disk, this may tilt the pre-existing disk to another rotation
axis. In fact, \citet{2011MNRAS.417.1817T} propose such a scenario to explain
misaligned exoplanets. \revised{More recently, the exchange of disk material and angular
momentum between two passing stars with disks has been studied as an alternative
way of tilting the disk and the resulting planetary system produced by it
\citep{2017MNRAS.471.2334X}.}

As a result of this angular momentum, cloudlet capture is thus a bit different
from the usual BHL accretion: while BHL accretion is the accretion of matter
directly onto the star, cloudlet capture, by virtue of the high angular momentum,
typically leads to the formation (or replenishment) of a circumstellar disk.
This material is bound to, but has not yet accreted onto the star.
Angular momentum redistribution within the disk (either due to viscous
disk accretion or gravitational instability) is then required to allow that
mass to find its way to the stellar surface.

\revised{If a protoplanetary disk already exists before the cloudlet capture
  event sets in, then the infalling material interacts in a complicated way with
  the pre-existing disk. As already alluded to above, it can tilt the disk. But
  as has been shown by \citet{2015A&A...582L...9L} and
  \citet{2017A&A...599A..86H}, such asymmetric infall can also drive 
  spiral waves in the disk, which can transport angular momentum and cause
  inward motion of the gas. In their models, for the case of asymmetric
  infall (corresponding to the asymmetric cloudlet capture in our terminology),
  even an $m=2$ mode is visible, which suggests a possible link to
  some of the observed $m=2$ spirals seen in
  scattered light \citep[e.g.][]{2017A&A...597A..42B, 2015A&A...578L...6B} and
  at millimeter-wavelengths \citep[e.g.][]{2016Sci...353.1519P,
    2018ApJ...869L..43H}. Furthermore, as
  shown by \citet{2015ApJ...805...15B}, infall onto the disk can lead
  to the formation of vortices which, in addition to being potential sites of
  planet formation, also appear to have their observational counterparts
  \citep[e.g.][]{2013Sci...340.1199V, 2012ApJ...754L..31C}.}

As we will show below, such off-center cloudlet capture events will often be
accompanied by arc-shaped nebulosity, because the part of the cloudlet that is
not accreted flies by in a hyperbolic orbit. Such ``arcs'' are, in fact, seen
around several Transitional Disk stars and FU Orionis stars, as we will
discuss in Section \ref{sec-obs}. 

If the cloudlet is substantially larger than $R_{\mathrm{HL}}$, then it is still
possible that the BHL accretion leads to a disk, as long as there is a
sufficiently large density gradient in the cloudlet as the star passes through
\citep{2005ApJ...618..757K}. For even larger cloudlets, however, the process
would approach the classical BHL accretion \citep{2006ApJ...638..369K}.

We will show that for cases in which $R_{\mathrm{cloud}}\gg b$, instead of
arcs, Bondi-Hoyle-like ``tails'' may be visible as reflection nebulosity.
Such tails also appear to be seen, for instance in Z CMa
\citep{2016SciA....200875L} and SU Aurigae \citep{2019AJ....157..165A}.

Since arc-shaped and tail-shaped reflection nebulosity appears to be found
around several Transitional Disks and FU Orionis stars, it is tempting to
speculate if the special properties of such sources may have their origin in
cloudlet capture events in the not-too-distant past. Transitional disks are a
special kind of protoplanetary disks which feature a large cavity in their inner
regions. This often leads to a geometry featuring an inner-disk, a large gap and
an outer disk. In some cases, the inner and outer disks appear to have wildly
different rotation axes: for the stars HD 142527 and HD 100453 these are of
order $\sim 70^\circ$ inclined with respect to each other
\citep{2015ApJ...798L..44M,2017A&A...597A..42B}. In this paper we speculate
whether the outer disks may have originated as a result of a cloudlet capture
event. FU Orionis stars are stars that undergo a large accretion outburst.
Given that several of these FU Orionis stars have arc/tail-shaped nebulosity
nearby \citep[e.g.][]{2016SciA....200875L}, we speculate that such outbursts may
be triggered by a recent cloudlet capture event.

This paper is structured as follows: We will first make some analytic estimates
in Section \ref{sec-estimates}. In Section \ref{sec-model} we will show some
results of simple hydrodynamic modeling.  We will then discuss observations of
reflection nebula patterns around a selected set of stars in Section
\ref{sec-obs}. In Section \ref{sec-discussion} we will discuss the
interpretation of these patterns in terms of the cloudlet capture and cloudlet flyby
scenario, and whether there may be a link to Transition Disks and FU Orionis
stars. We will also discuss the limitations of the model.

\section{Estimations}
\label{sec-estimates}
Before we resort to numerical hydrodynamic calculations, we make a few simple
estimations of the process of cloudlet capture and cloudlet flyby. For simplicity we
assume the cloudlet to be spherical, with radius $R_{\mathrm{cloud}}$, constant
density $\rho_{\mathrm{cloud}}$ and temperature $T_{\mathrm{cloud}}$,
approaching the star with a velocity at infinity $v_{\infty}$ and impact
parameter $b$. The cloudlet mass is much smaller than the stellar mass, and we
assume that the cloudlet is in pressure equilibrium with a low-density warm neutral
medium of $T_{\mathrm{wnm}}=8000$ K \citep{1969ApJ...155L.149F}. The motion of
the cloud, as it approaches the star, will thus initially be largely ballistic.

\subsection{Critical impact parameter}
A test particle approaching the star with impact parameter $b$ and velocity
$v_\infty$ will follow a hyperbolic orbit with a deflection angle of
\begin{equation}
\theta_{\mathrm{deflect}} = 2\,\mathrm{arcsin}\left(\frac{1}{e}\right)
\end{equation}
where the eccentricity $e$ is defined as
\begin{equation}
e = \sqrt{1+\frac{b^2}{b_{\mathrm{crit}}^2}}
\end{equation}
and the critical impact parameter $b_{\mathrm{crit}}$ is defined as
\begin{equation}\label{eq-bcrit-def}
b_{\mathrm{crit}} = \frac{GM_{*}}{v_{\infty}^2} = \tfrac{1}{2}R_{\mathrm{HL}}
\end{equation}
The closest approach occurs at a distance
$r_{\mathrm{close}}=b\sqrt{(e-1)/(e+1)}$. The meaning of $b_{\mathrm{crit}}$ is
the impact parameter for which the deflection angle is 90$^\circ$.  For
$b=b_{\mathrm{crit}}$ the closest approach occurs at
$0.41\,b_{\mathrm{crit}}$. For a finite-size cloudlet we thus expect that an
encounter with $b\simeq b_{\mathrm{crit}}$ will yield a well-defined arc, as the
cloudlet will be tidally stretched roughly along the hyperbolic orbit. If the cloudlet
has a radius $R_{\mathrm{cloud}}\gtrsim b_{\mathrm{crit}}$, some material from
the cloudlet will get captured and forms a disk, while the rest of the cloudlet will
fly by and forms the arc.

Given typical random velocities of filaments and young stars in giant molecular
clouds of $\sim 1$ km/s, and taking the typical stellar mass of a Herbig Ae
star of $M_{*}=2.5\,M_{\odot}$, Eq.~(\ref{eq-bcrit-def}) with
$v_{\infty}=1\,\mathrm{km}/\mathrm{s}$ yields $b_{\mathrm{crit}}\simeq 2200$
au. This means that arc-shaped nebulosity around Herbig Ae stars, if observed,
is expected to have spatial scales of the order of a few thousand au, and last a
Kepler time scale of about $10^5$ years. According to
\citet{2010A&A...520A..17K}, however, this estimate of
$v_{\infty}=1\,\mathrm{km}/\mathrm{s}$ may be a bit on the high side. If we take
instead $v_{\infty}=0.5\,\mathrm{km}/\mathrm{s}$ we obtain
$b_{\mathrm{crit}}\simeq 10^{4}$ au, and a time scale of half a million years.

\subsection{Mass and size of cloudlets}
\citet{2010A&A...520A..17K} give a simple formula (their Eq.~19) that relates
the cloud/cloudlet mass to its size scale, based on observational data of GMCs from
\citet{2004Ap&SS.292...89F}. We rewrite that formula as:
\begin{equation}\label{eq-mass-radius-relation}
M_{\mathrm{cloud}} \simeq
0.01\,M_{\odot}\,\left(\frac{R_{\mathrm{cloud}}}{5000\,\mathrm{au}}\right)^{2.3}
\end{equation}
where we replaced the length scale $L$ in \citet{2010A&A...520A..17K} by
$2R_{\mathrm{cloud}}$. With a mean molecular weight of 2.3 for a gas consisting
of molecular hydrogen and helium, this leads to a gas number density of
\begin{equation}\label{eq-estim-cloud-density}
n_{\mathrm{cloud}} \simeq 2.8\times 10^3\;\left(\frac{M_{\mathrm{cloud}}}{0.01\,M_{\odot}}\right)^{-0.30}\;\mathrm{cm}^{-3}
\end{equation}
There is, of course, some scatter in this relation, and it may vary somewhat
between different GMCs. We will, however, use this relation for our model
setup. Note that if we assume $R_{\mathrm{cloud}} \simeq b_{\mathrm{crit}}$, in
which case we expect some of the cloudlet to be captured, and the other part to
produce an strongly bent arc, then we find that $M_{\mathrm{cloud}}\propto
v_{\infty}^{-4.6}$, which is a very steep dependency. Only a small variation in
$v_{\infty}$ could lead to vastly different cloudlet masses under this
assumption. For $v_{\infty}=1\,\mathrm{km/s}$ we find $M_{\mathrm{cloud}}\simeq
1.5\times 10^{-3}\,M_{\odot}$, while for $v_{\infty}=0.5\,\mathrm{km/s}$ we
obtain $M_{\mathrm{cloud}}\simeq 3.5\times 10^{-2}\,M_{\odot}$. Note that all
these values assume a stellar mass of 2.5 $M_{\odot}$, appropriate for a Herbig
Ae star.

\section{Hydrodynamic models}
\label{sec-model}
\subsection{Numerical hydrodynamic model setup}
Using the PLUTO hydrodynamics code\footnote{\url{http://plutocode.ph.unito.it}
  version 4.1} \citep{2007ApJS..170..228M}
we now demonstrate that the process of cloudlet capture is often
associated with the formation of an arc-shaped reflection nebula like the ones
seens around some Transition Disks and FU Orionis stars. We will focus on these
large-scale features, and leave the detailed study of the formation and/or
feeding of the disk to a \revised{follow-up paper (K\"uffmeier,
  Goicovic \& Dullemond in prep)}.

For our model we choose the same simple scenario of a spherical cloudlet
approaching the $M_{*}=2.5\,M_{\odot}$ star as in Section
\ref{sec-estimates}. Given the radius of the cloudlet $R_{\mathrm{cloud}}$, the
cloudlet gas density is given by Eq.~(\ref{eq-estim-cloud-density}). We place this
cloudlet at the start of the simulation at a distance from the star
substantially larger than $b_{\mathrm{crit}}$ (Eq.~\ref{eq-bcrit-def}), so that
the initial movement of the cloudlet will be nearly linear. We set
$v_{\infty}=1\,\mathrm{km}/\mathrm{s}$. The critical impact parameter is
then $b_{\mathrm{crit}}=2200\,\mathrm{au}$.

We do two sets of models: adiabatic models and isothermal models, representing
two extreme cases: that of no cooling and that of instant thermal adaption to
the environment. We do not include time-dependent heating/cooling in the models.
\revised{We also do not include magnetic fields, although they likely play a
role.}

For the adiabatic model we choose $T_{\mathrm{cloud}}=30\,\mathrm{K}$ as the
cloudlet temperature. The cloudlet must be embedded in a warm neutral medium in order
to be kept under pressure, otherwise it will thermally expand well before it
reaches the star. This warm neutral medium is set at a temperature of 8000 K. By
demanding pressure equilibrium, the density of this medium is set. The ratio of
specific heats is $\gamma=5/3$ for both the cold cloudlet (because molecular
hydrogen is too cold to excite rotational levels) and for the warm neutral
medium (because the hydrogen will be atomic). One problem with the adiabatic
model assumption is that material that gets captured inside the potential well
is likely to be shock-compressed and hot. This prevents the formation of a disk,
and it will cause most of the captured material to ``bounce back'' and escape
again in a wide range of directions. The isothermal models do not have this
problem. A disk can be readily formed. But for the isothermal model it is
impossible to embed the cloudlet into a confining medium (it would make the model
non-isothermal). \revised{In principle one could make an initial condition
  consisting of two isothermal states: a cold isothermal cloudlet inside a
  hot isothermal medium, as we do for the adiabatic case. But some mixing
  between the two phases will occur due to numerical diffusion, at which
  point it will no longer be possible to decide which of the two temperatures
  to take.} The cloudlet expansion problem can \revised{thus} not be avoided
\revised{for the isothermal models}, and a
cloudlet can therefore not really travel very far before it thermally expands again
and dissipates. In a supersonically turbulent star formation environment such
short-lived cloudlets can form through colliding flows. \revised{In fact,
  3-D simulations of turbulent molecular cloud complexes show that such
  transient cloudlets are formed (and dissipated) all the time
  \citep{2004RvMP...76..125M}.} In our isothermal models
we therefore put the cloudlet initially already much closer to the star than in the
adiabatic models, and we set the temperature to
$T_{\mathrm{cloud}}=10\,\mathrm{K}$. In this way the cloudlet can get captured
before it dissipates.

The adiabatic models are set up in cartesian coordinates, in a flattened box
with $x\in [-2L, L]$, $y\in [-L,L]$ and $z\in [-L/4,L/4]$, where
$L=4.6\,R_{\mathrm{cloud}}$ is chosen such that the cloudlet fits vertically inside
the box. The cloudlet is initially placed at $x=-1.7\,L$, $y=-b$ and $z=0$, or in
vector form $\mathbf{x}\equiv (x,y,z)=(-1.7L,-b,0)^T$. For the isothermal models
the box size is $x\in [-L, L]$, $y\in [-L,L]$ and $z\in [-L/4,L/4]$ and
the initial position is at $\mathbf{x}\equiv (x,y,z)=(-0.7L,-b,0)^T$.
The initial velocity of
the gas is taken to be constant throughout the grid, with velocity vector
$\mathbf{v}=(1,0,0)\,\mathrm{km/s}$. The grid is composed of $384 \times 256
\times 64$ grid cells for the adiabatic models and $256 \times 256
\times 64$ grid cells for the isothermal models.
The star is located at $\mathbf{x}=(0,0,0)$. Given the
mirror symmetry in the z-plane, we only model the upper half, and put a mirror
symmetry boundary condition at $z=0$. The conditions at the other boundaries are
simply copies of the warm neutral medium hydrodynamic state in the ghost
cells. This allows waves to flow off the grid without much reflection. The
gravitational potential is smoothed near the origin as
$\Phi=-GM_{*}/(r^8+r_{\mathrm{sm}}^8)^{1/8}$, with smoothing radius
$r_{\mathrm{sm}}=0.013\,L$. The model does not include self-gravity, nor a
gravitational back-reaction onto the star. We run the model for 3 crossing times
along the $x$-axis and make 61 dumps in equal time intervals. PLUTO works in
dimensionless code units; we choose a length unit of $100\,\mathrm{au}$, a
velocity unit of $1\,\mathrm{km/s}$ and a density unit of $10^3\,m_p$, with
$m_p$ being the proton mass. But given the hydrodynamic nature of this model
setup, the results are scalable.

The only remaining dimensionless \revised{physical} free parameters of this setup are
$R_{\mathrm{cloud}}/b_{\mathrm{crit}}$ and $b/b_{\mathrm{crit}}$.

\subsection{Postprocessing radiative transfer for scattered light images}
The appearance of the cloudlet, as it flies by and/or partly gets captured, can be
estimated using a simple radiative transfer setup. Rather than applying a
fully-fledged Monte Carlo radiative transfer code such as RADMC-3D, we make use
of the fact that these cloudlets are typically optically thin to stellar
radiation. The scattering source function at each
location ${\bf x}$ is then 
\begin{equation}
j_\nu^{\mathrm{scat}}(\mathbf{x}) = \frac{F_\nu^{*}(\mathbf{x})}{4\pi}\rho_d(\mathbf{x})\kappa_\nu^{\mathrm{scat}}\varphi(\theta(\mathbf{x}))
\end{equation}
where $F_\nu^{*}$ is the stellar flux as seen at $\mathbf{x}$
\begin{equation}
F_\nu^{*} = \frac{L_\nu^{*}}{4\pi r^2}
\end{equation}
with $r\equiv |\mathbf{x}|$ and $L_\nu^{*}$ is the stellar luminosity at
frequency $\nu$. The dust density $\rho_d(\mathbf{x})$ is taken to be 0.01 times
the gas density $\rho_g(\mathbf{x})$ and the dust scattering opacity
$\kappa_\nu^{\mathrm{scat}}$ is computed using Mie theory for spherical dust
grains with a Gaussian size distribution centered around a radius of $a=1\,
\mu\mathrm{m}$ and a half-width in $\lg(a)$ of 0.05, made of pyroxene with 70\%
magnesium with a material density of 3 g/cm$^3$. The function $\varphi(\theta)$
is the scattering phase function, where $\theta(\mathbf{x})$ is the angle
between the stellar radiation at position $\mathbf{x}$ and the observer. We put
the observer at $z=+\infty$, so that
\begin{equation}
\cos(\theta(\mathbf{x})) = \frac{\mathbf{x}\cdot \mathbf{e}_z}{|\mathbf{x}|}
\end{equation}
with $\mathbf{e}_z=(0,0,1)^T$. The phase function is normalized such that for
isotropic scattering one would have $\varphi(\theta)=1$. We use the Henyey-Greenstein phase function
with $g$ computed with the Mie algorithm from $\langle\cos(\theta)\rangle$.
After computing $j_\nu^{\mathrm{scat}}(\mathbf{x})$ in each cell in the grid,
we numerically integrate the formal radiative transfer equation from $z=-\infty$
to $z=+\infty$ to obtain the scattered light image:
\begin{equation}
I_\nu^{\mathrm{obs}}(x,y) = \int_{-\infty}^{+\infty} j_\nu^{\mathrm{scat}}(x,y,z) dz
\end{equation}
where the contribution is only non-zero within the grid spanning between
$z=-L/4$ to $z=+L/4$. We carry out this computation at a wavelength of
$\lambda=0.65\,\mu\mathrm{m}$. The stellar radius is taken to be
$R_{*}=2.4\,R_{\odot}$ and the effective temperature of the star is
$T_{*}=10^4\,\mathrm{K}$. For simplicity we assume a Planck spectrum, but one
could also take e.g.\ a Kurucz spectrum. For our choice we obtain
$L_\nu=1.96\times 10^{20}\,\mathrm{erg}\,\mathrm{s}^{-1}\,\mathrm{Hz}^{-1}$.
The scattering opacity at this wavelength is
$\kappa_{\nu}^{\mathrm{sca}}=6.0\times 10^3\,\mathrm{cm}^2/\mathrm{g}$, which
is cross section per gram of dust; we assume a dust-to-gas ratio of $0.01$.
The Henyey-Greenstein phase parameter at this wavelength is $g=0.73$.

\subsection{Results of the adiabatic models}
\begin{table*}
  \centerline{\begin{tabular}{l|ccccccc}
   & EOS &  $R_{\mathrm{cloud}}$ [au] & $b$ [au] & $v_\infty$ [km/s] & $T_{\mathrm{cloud}}$ [K] & $M_{*}/M_{\odot}$ & $b_{\mathrm{crit}}$ [au]\\
  \hline
  A1 & adiabatic &  887 & \revised{1774} & 1.0 & 30 & 2.5 & 2218\\
  A2 & adiabatic & 1330 & \revised{1774} & 1.0 & 30 & 2.5 & 2218\\
  A3 & adiabatic & 2662 & 2218 & 1.0 & 30 & 2.5 & 2218\\
  \hline
  I1 & isothermal &  887 & \revised{1774} & 1.0 & \revised{10} & 2.5 & 2218\\
  I2 & isothermal & 1330 & \revised{1774} & 1.0 & \revised{10} & 2.5 & 2218\\
  I3 & isothermal & 2662 & 2218 & 1.0 & \revised{10} & 2.5 & 2218\\
  \end{tabular}}
\caption{\label{table-model-params}Overview of the model parameters.}
\end{table*}
\revised{The parameters of the adiabatic models A1, A2 and A3 are listed in
  Table \ref{table-model-params}. All models have approach-velocity
  $v_{\infty}=1\,\mathrm{km/s}$, cloudlet temperature
  $T_{\mathrm{cloud}}=30\,\mathrm{K}$, and stellar mass $M_{*}=2.5\,M_{\odot}$.}

\begin{figure*}[h]
  \centerline{
    \includegraphics[width=0.48\textwidth]{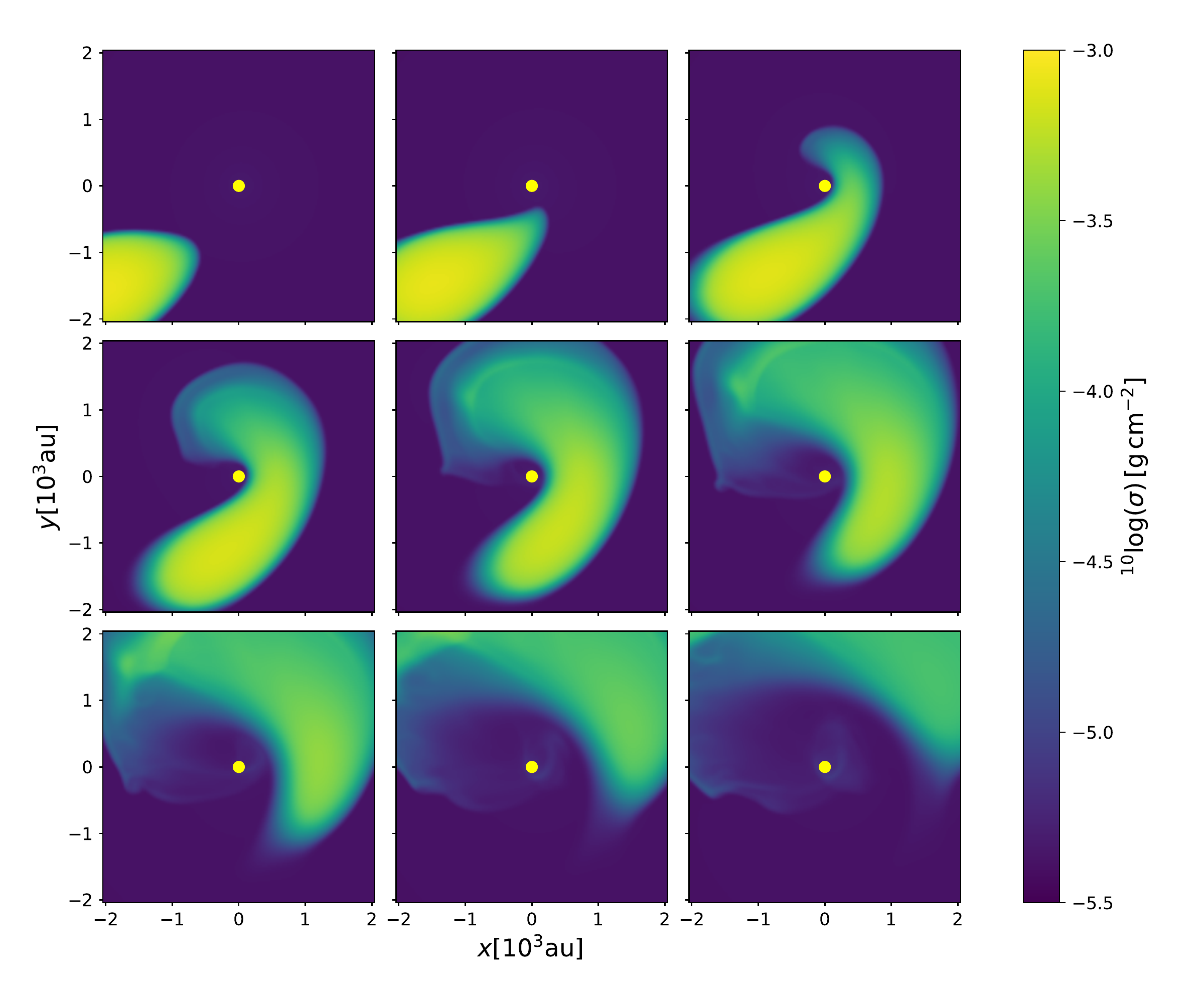}
    \includegraphics[width=0.48\textwidth]{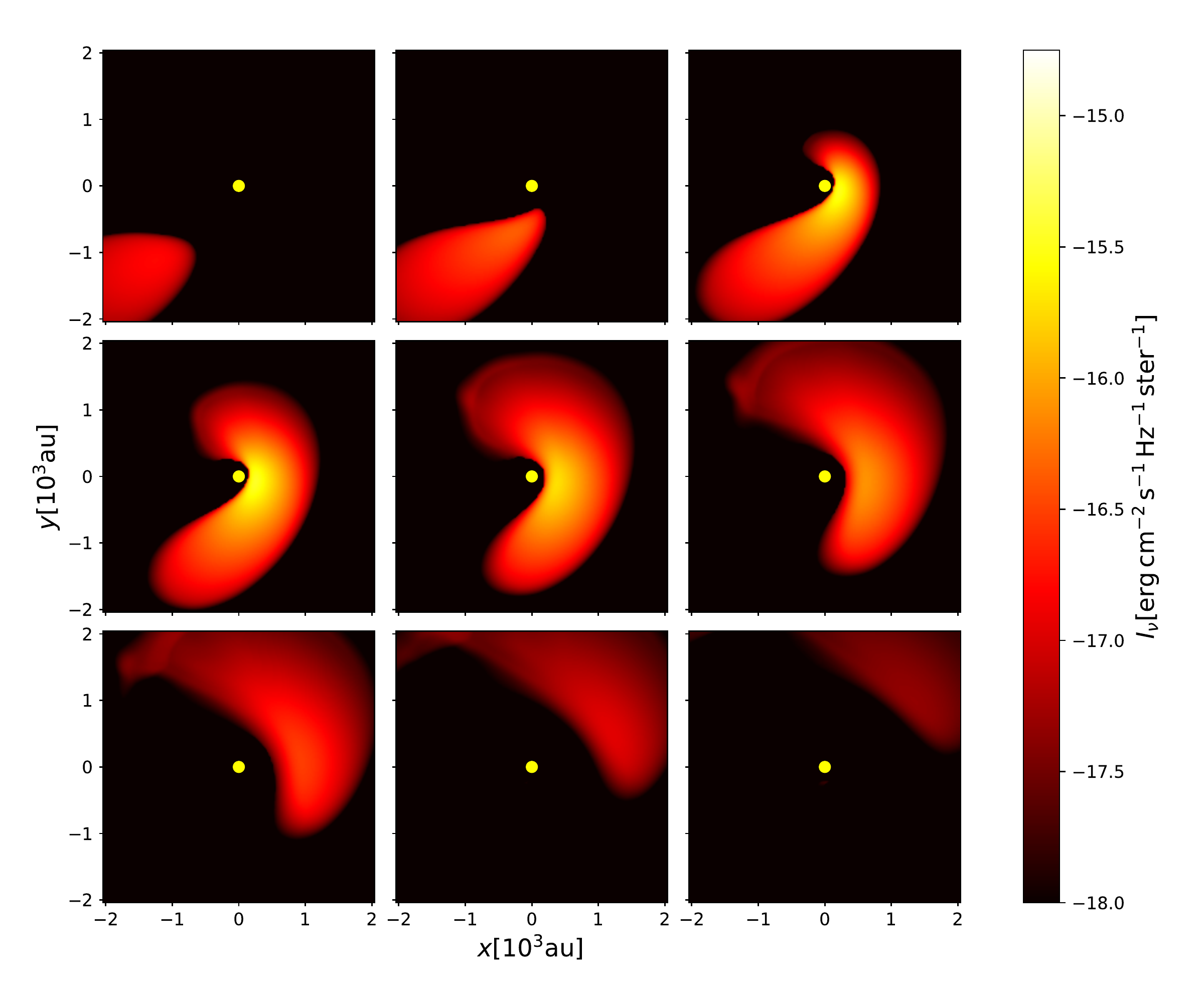}
  }
  \caption{\label{fig-model-A1-scat9}\revised{Snapshots of model A1}
    with $R_{\mathrm{cloud}}=0.4\,b_{\mathrm{crit}}=887\,\mathrm{au}$ and
    $b=0.8\,b_{\mathrm{crit}}=1774\,\mathrm{au}$. \revised{Left panel:
      column density, right panel: synthetic scattered light image at
      $\lambda=0.65\mu\mathrm{m}$. In each panel, the 3$\times$3 subpanels
    are different times, where} time goes from top-left to bottom-right with intervals
    of $1934\,\mathrm{yr}$. The yellow dot marks the
    location of the star. Note that the model domain extends well beyond the
    field of view shown here.}
\end{figure*}

In Fig.~\ref{fig-model-A1-scat9} a time sequence of \revised{column density
and} synthetic scattered light
images is shown for adiabatic model A1, which has
$R_{\mathrm{cloud}}=0.4\,b_{\mathrm{crit}}=887\,\mathrm{au}$ and
$b=0.8\,b_{\mathrm{crit}}=1774\,\mathrm{au}$.
This is a model in which the cloudlet impact parameter is small enough to cause a
substantial gravitational deflection of the orbit of the cloudlet, but the cloudlet is
itself too small to be partially captured by the star. It results in an curved
flyby and a clear arc-shaped reflection nebula. The vast majority of the mass of
the cloudlet flies by, but one can still notice a tiny bit of reflection much
closer to the star, indicating that not all of the cloudlet mass avoided capture.

\begin{figure*}[h]
  \centerline{
    \includegraphics[width=0.48\textwidth]{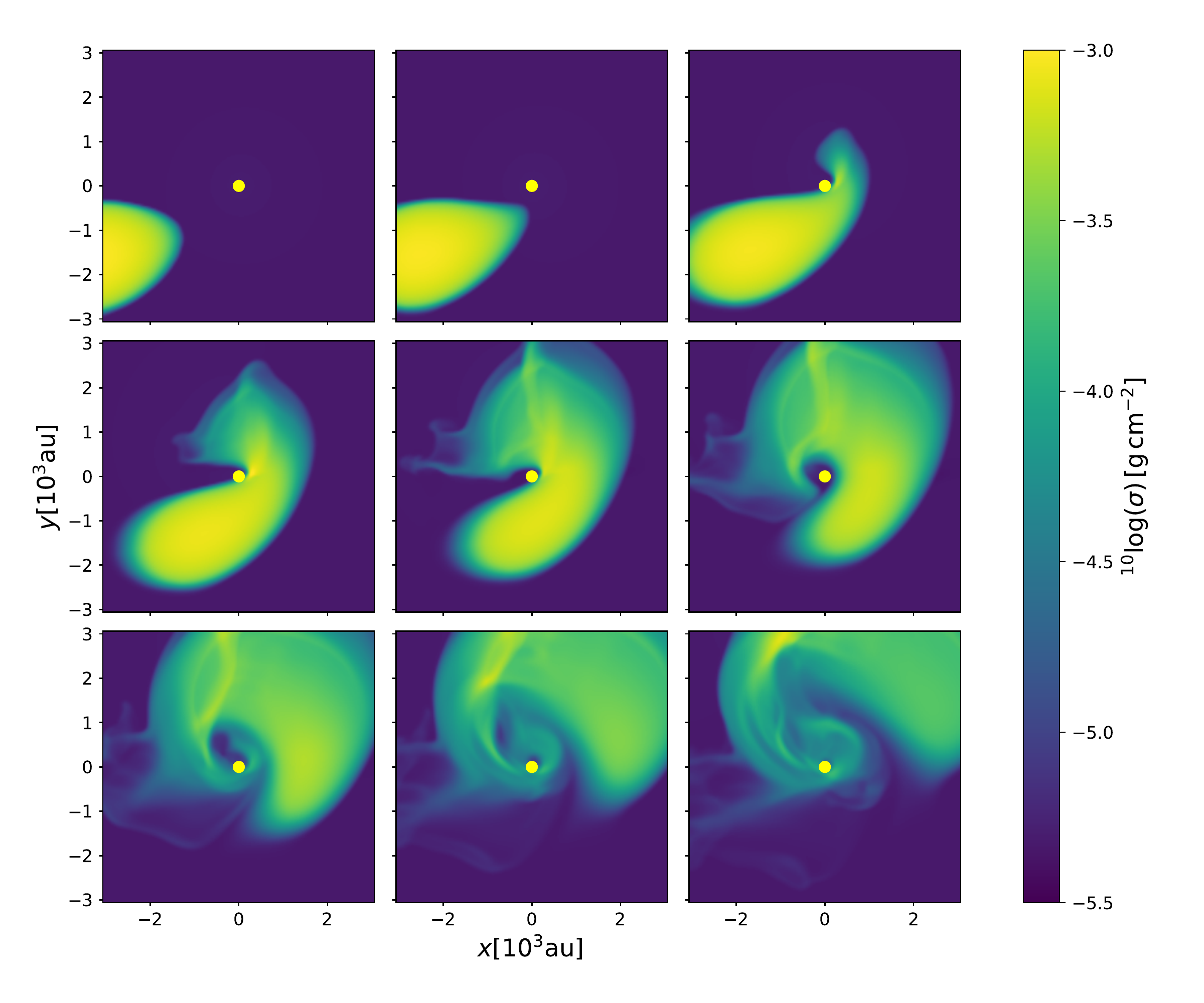}
    \includegraphics[width=0.48\textwidth]{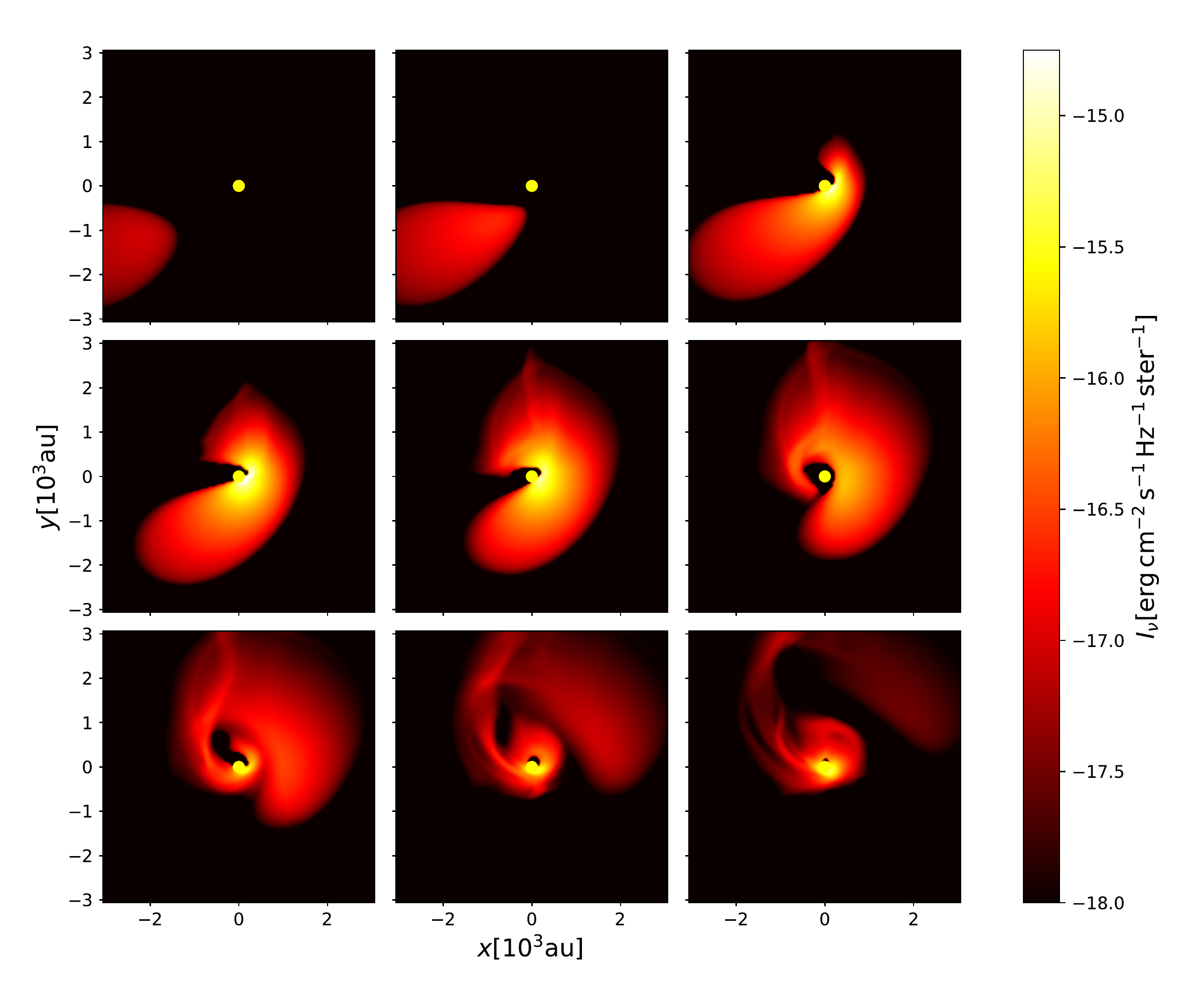}
  }
  \caption{\label{fig-model-A2-scat9}As Fig.~\ref{fig-model-A1-scat9} but now for model
    A2 which has a larger cloudlet size. \revised{Note the different axis
  size compared to Fig.~\ref{fig-model-A1-scat9}.}}
\end{figure*}

In Fig.~\ref{fig-model-A2-scat9} the same images are shown for
adiabatic model A2, which has a larger cloudlet radius
($R_{\mathrm{cloud}}=0.6\,b_{\mathrm{crit}}=1330\,\mathrm{au}$) but still the
same impact parameter as in model A1. In this case the cloudlet is sufficiently
large that a \revised{non-negligible} amount of material is captured and remains bound, while
the remainder of the cloudlet flies by. Hence, during the closest approach of the
cloudlet, both an arc is seen as well as freshly captured circumstellar material.
\revised{Still, most of the material avoids capture and escapes.}

\begin{figure*}[h]
  \centerline{
    \includegraphics[width=0.48\textwidth]{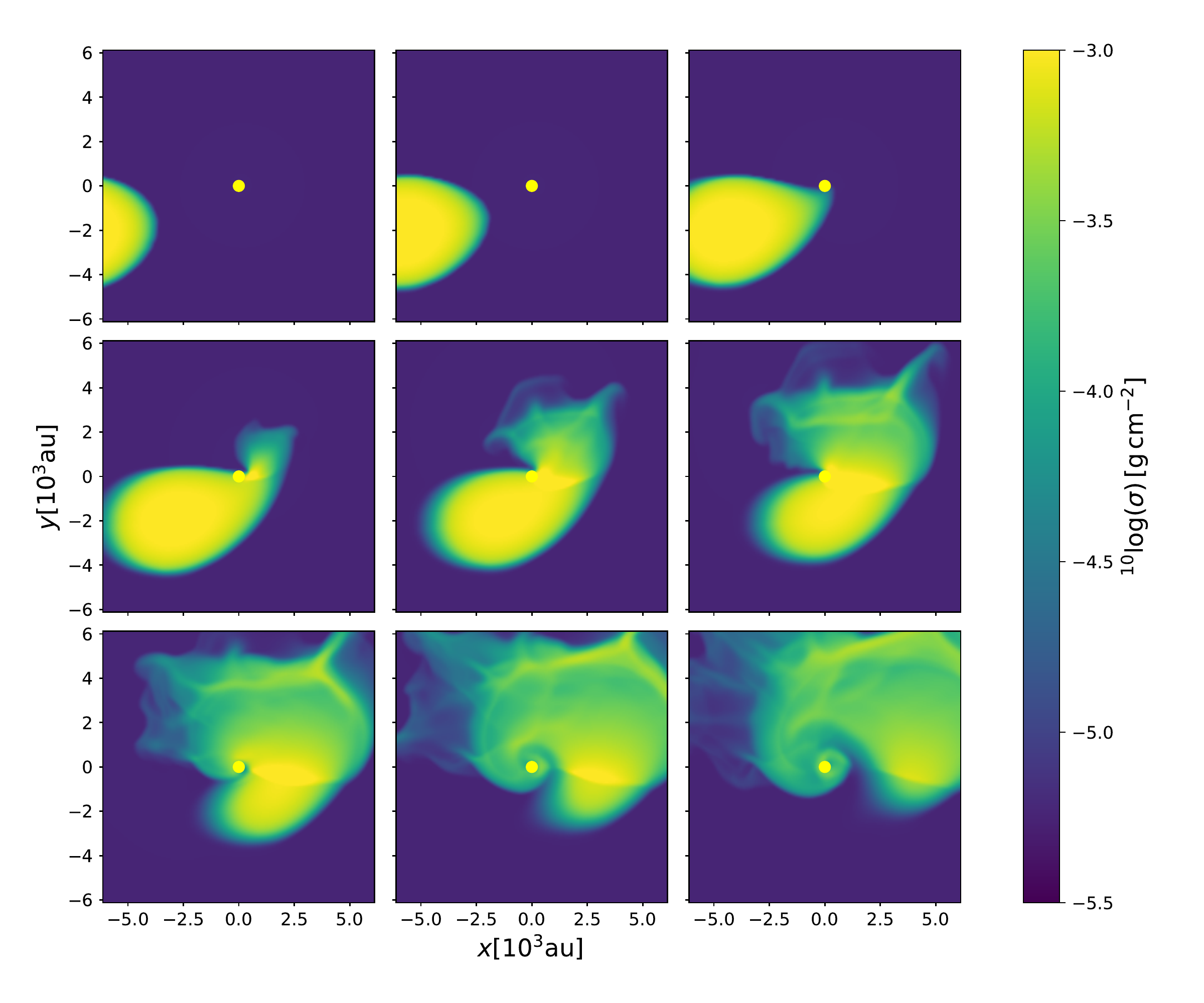}
    \includegraphics[width=0.48\textwidth]{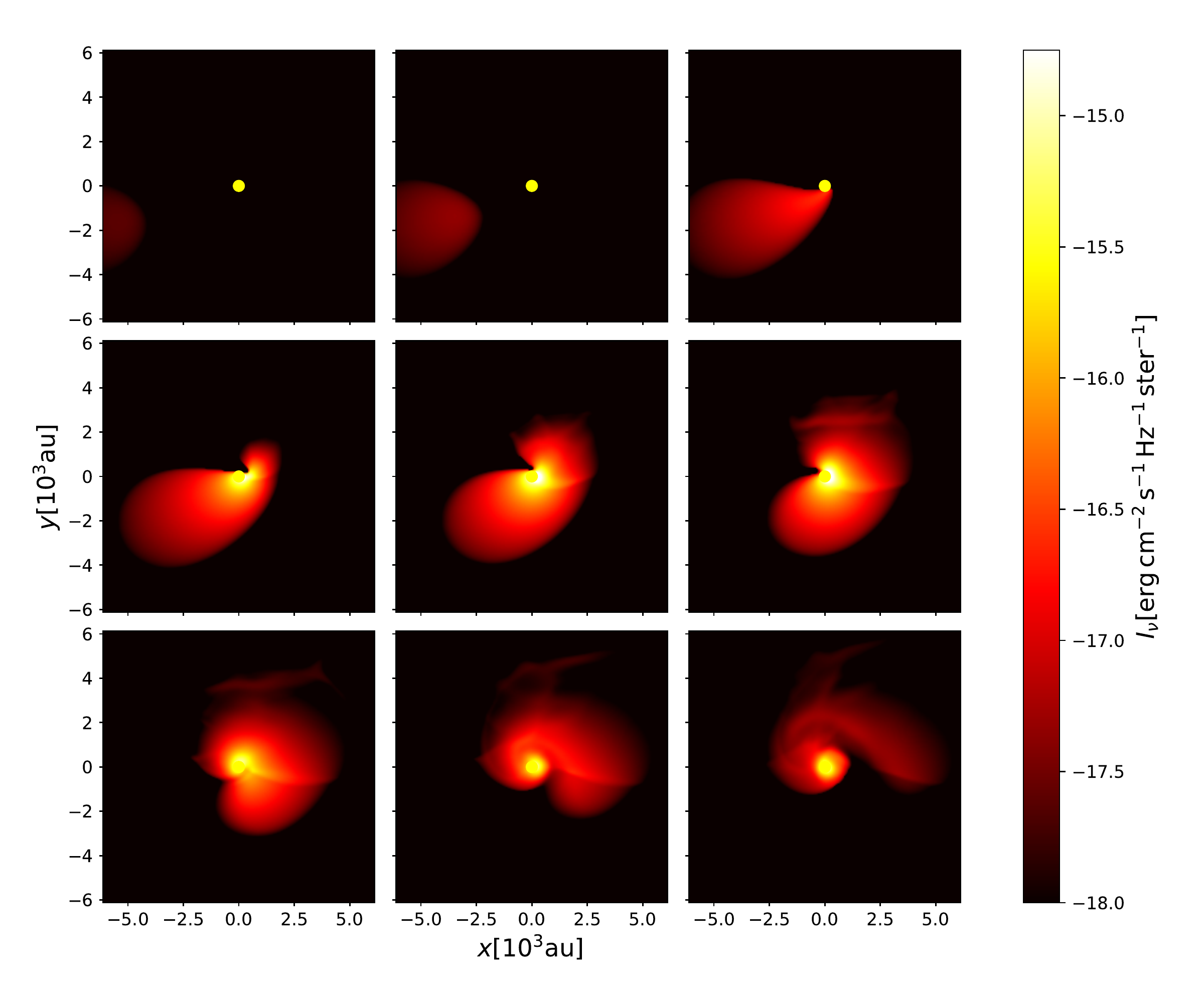}
  }
  \caption{\label{fig-model-A3-scat9}As Fig.~\ref{fig-model-A1-scat9} but now for model
    A3 which has $R_{\mathrm{cloud}}=1.2\,b_{\mathrm{crit}}=2662\,\mathrm{au}$ and
    $b=1.0\,b_{\mathrm{crit}}=2218\,\mathrm{au}$. \revised{Note the different axis
  size compared to Fig.~\ref{fig-model-A1-scat9}.}}
\end{figure*}

Finally in Fig.~\ref{fig-model-A3-scat9} a time sequence
is shown for adiabatic model A3, which has
$R_{\mathrm{cloud}}=1.2\,b_{\mathrm{crit}}=2662\,\mathrm{au}$ and
$b=1.0\,b_{\mathrm{crit}}=2218\,\mathrm{au}$. In this case an arc is 
seen, but also a Bondi-Hoyle type tail superposed on it. This is because not all
matter of the cloudlet passes by the star from one side. Some of the cloudlet
material has, so to speak, a negative impact parameter, and collides with the
majority of material coming from the positive impact parameter side.
\revised{As in the case of model A2, some material remains bound to
the star, forming a circumstellar disk.}

In all these models the encounter of the cloudlet with the star leads to
reflection nebulosity around the star of material that is mostly gravitationally
unbound to the star. The arc shapes are the result of the gravitational
deflection of the material that is passing by the star. The arc does not lie
exactly along the hyperbolic orbit: due to the size of the cloudlet being of
order $b_{\mathrm{crit}}$, the arc, being initially stretched more or less
along the orbit, eventually expands almost spherically away from the star.

\revised{Of the adiabatic models, the first one (A1) most clearly displays the
  arc-shaped nebulosity, and it behaves pretty much as one would expect on the
  basis of the ballistic arguments of Section \ref{sec-estimates}.  For A2, and
  even more so for A3, this simple picture fails, because the ballistic orbits
  cross each other, leading to the gas being shocked, and deflected from its
  ballistic path.}

\subsection{Results of the isothermal models}
Like for the adiabatic models the parameters are listed in Table
\ref{table-model-params}. \revised{Models I1, I2 and I3 share the cloudlet
  radius and impact parameter with the models A1, A2, and A3, respectively.
  They can thus, to some extent, be regarded as their isothermal counterparts,
  albeit at lower temperature ($T_{\mathrm{cloud}}=10\,\mathrm{K}$ instead of
  $T_{\mathrm{cloud}}=30\,\mathrm{K}$).}

\begin{figure*}[h]
  \centerline{
    \includegraphics[width=0.48\textwidth]{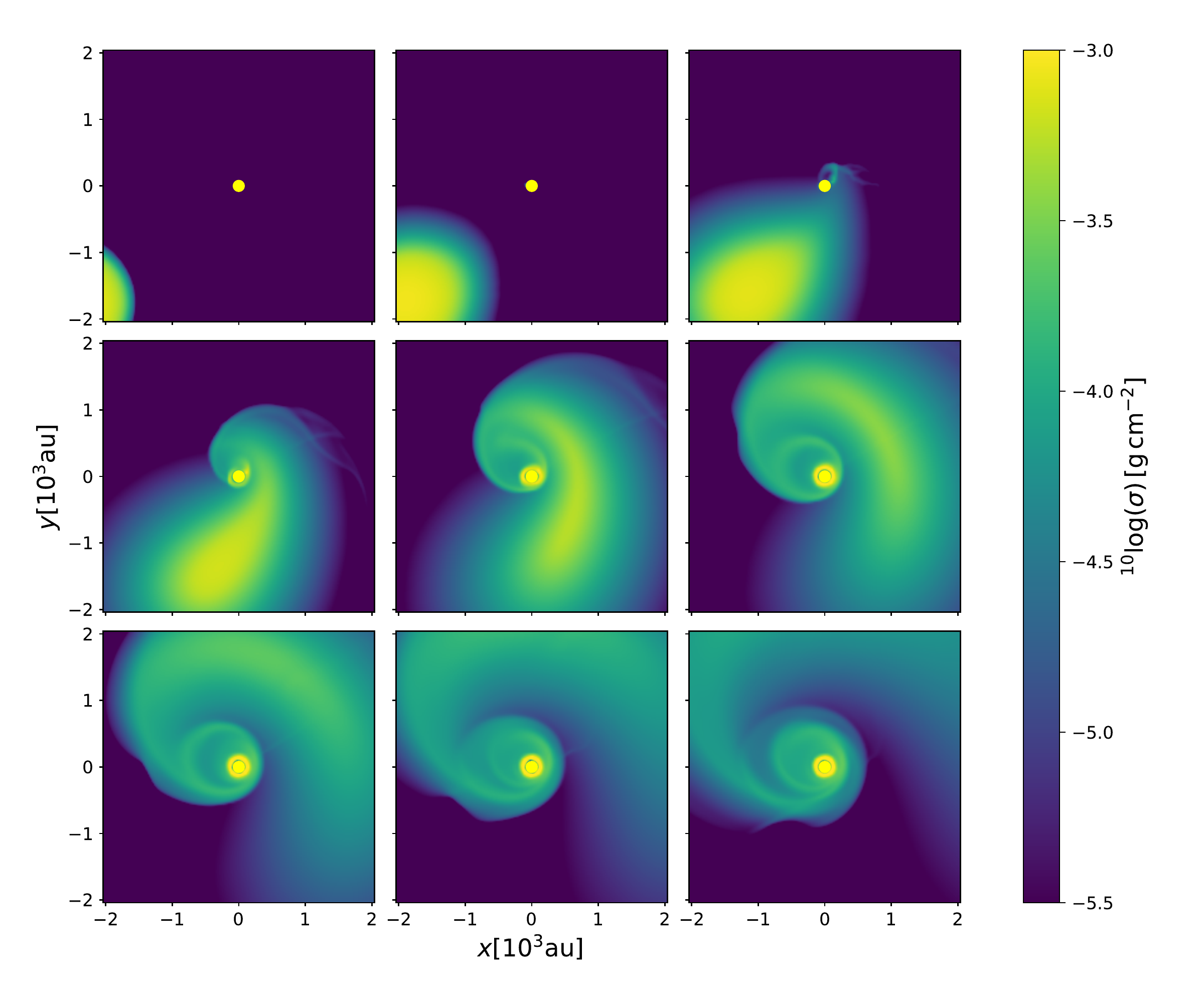}
    \includegraphics[width=0.48\textwidth]{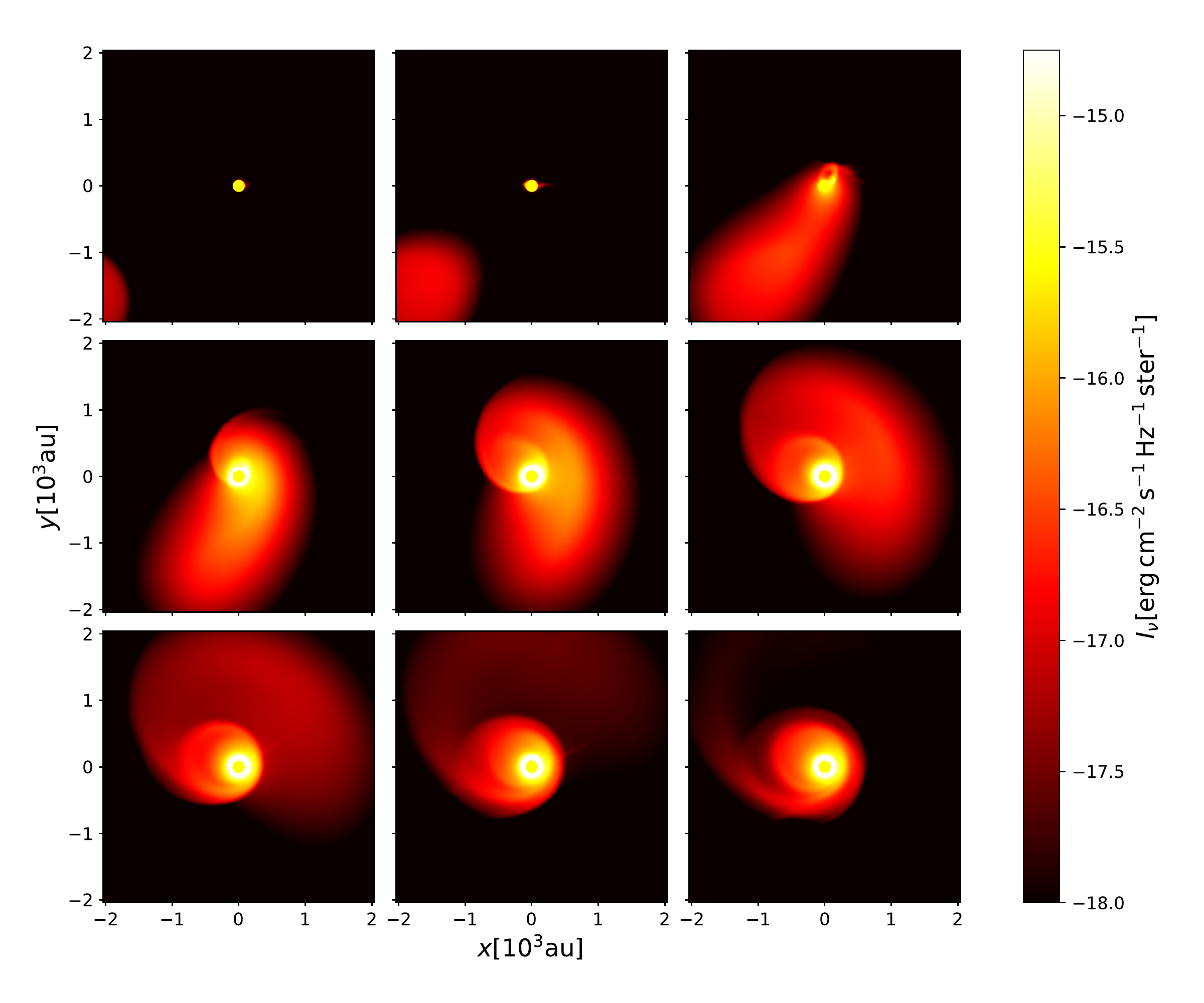}
  }
  \caption{\label{fig-model-I1-scat9}As Fig.~\ref{fig-model-A1-scat9} but now
    for model I1 which has an isothermal equation of state. The parameters are
    with $R_{\mathrm{cloud}}=0.4\,b_{\mathrm{crit}}=887\,\mathrm{au}$ and
    $b=0.8\,b_{\mathrm{crit}}=1774\,\mathrm{au}$. Left panel: column density,
    right panel: synthetic scattered light image at
    $\lambda=0.65\mu\mathrm{m}$. In each panel, the 3$\times$3 subpanels are
    different times, where time goes from top-left to bottom-right with
    intervals of $1934\,\mathrm{yr}$. The yellow dot marks the location of the
    star.}
\end{figure*}

For model I1 the scattered light images are shown in
Fig.~\ref{fig-model-I1-scat9}. The arc structure is visible, \revised{in
  particular in the 6th time snapshot (middle row, right),} but it is somewhat
smeared out. This is due to the thermal expansion of the cloudlet during the
capture. \revised{More prominent is the stretched ``arm'' seen in the third panel in
  the scattered light images (top row, right). This is the combined effect of
  tidal stetching of the accreting cloud and the $1/r^2$ dilution of the
  stellar light that scatters off the dust particles.}

\begin{figure*}[h]
  \centerline{
    \includegraphics[width=0.48\textwidth]{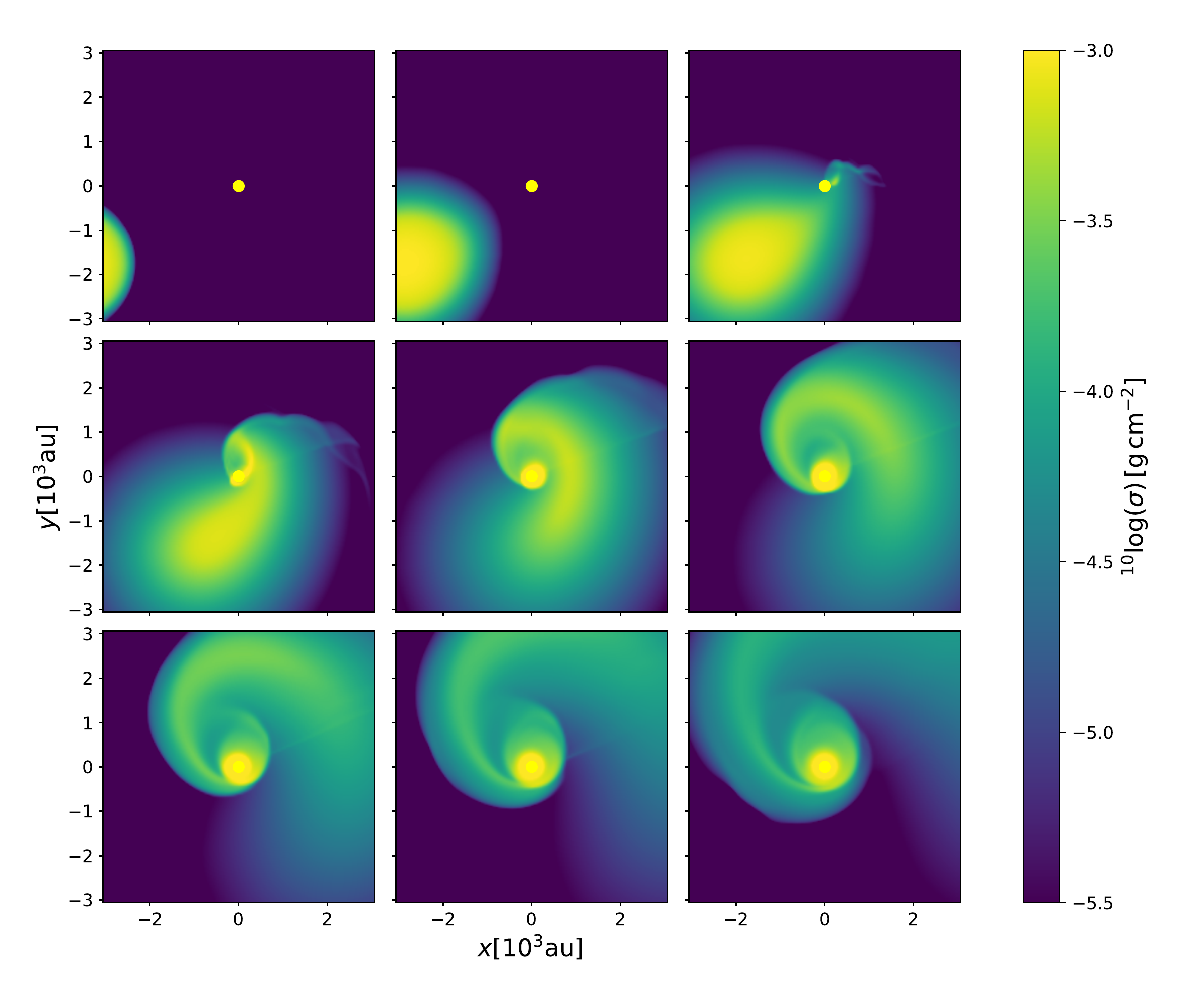}
    \includegraphics[width=0.48\textwidth]{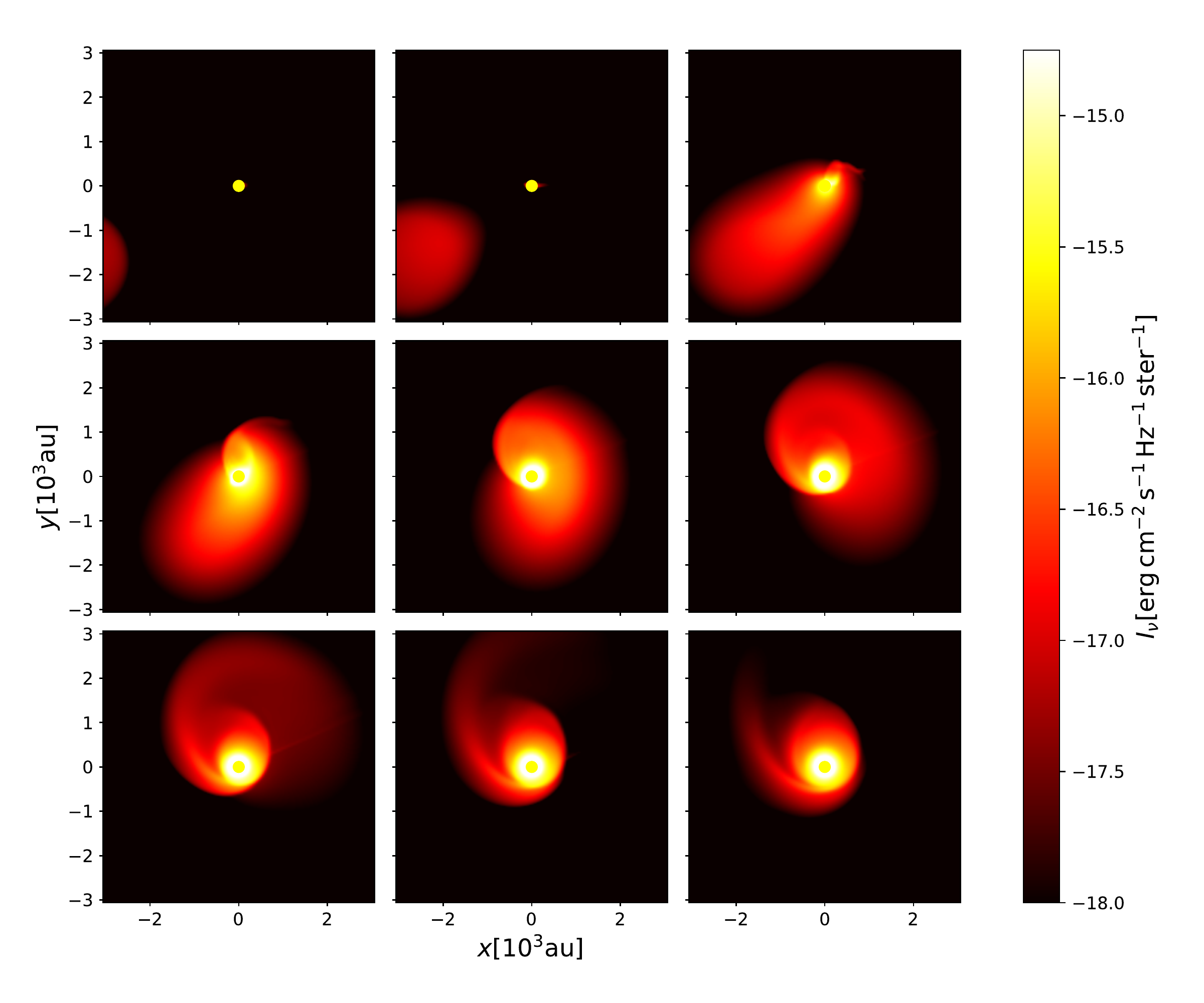}
  }
  \caption{\label{fig-model-I2-scat9}As Fig.~\ref{fig-model-I1-scat9}
    but now for model I2, which has a larger cloudlet size (i.e.~like model
    A2, but now isothermal). \revised{Note the different axis
  size compared to Fig.~\ref{fig-model-I1-scat9}.}}
\end{figure*}

\revised{In Fig.~\ref{fig-model-I2-scat9} the results are shown of model
  I2 in which, compared to model I1 the cloudlet is larger. The differences
  with model I1 are not nearly as prominent as between models A2 and A1. The
  arc shaped structure is, however, weaker than in model I1. In both models,
  however, the snapshots (the bottom row) show another prominent feature:
  an m=1 spiral arm. This is most clearly seen in the column density maps,
  but it is also visible in the scattered light images.}

\begin{figure*}[h]
  \centerline{
    \includegraphics[width=0.48\textwidth]{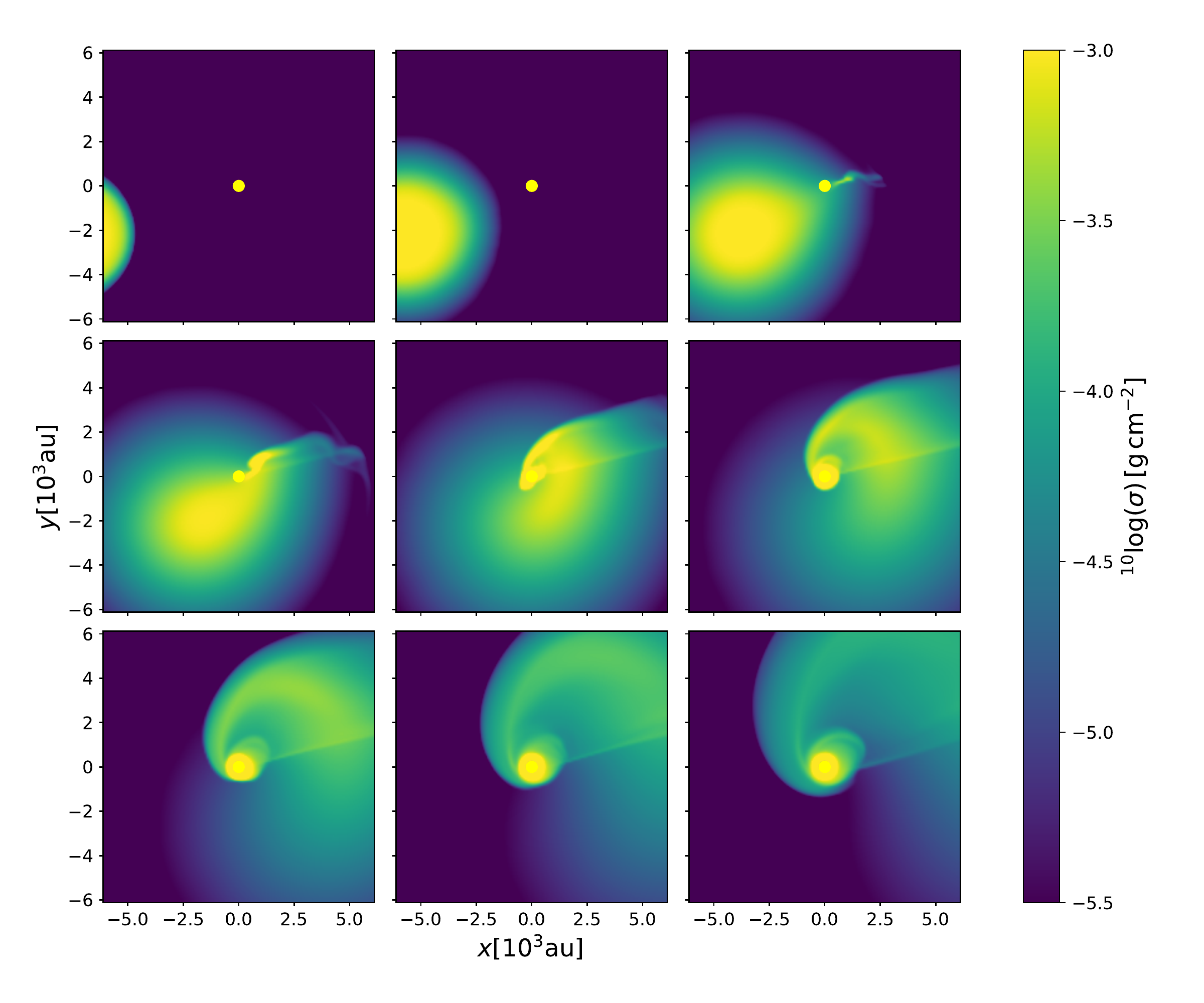}
    \includegraphics[width=0.48\textwidth]{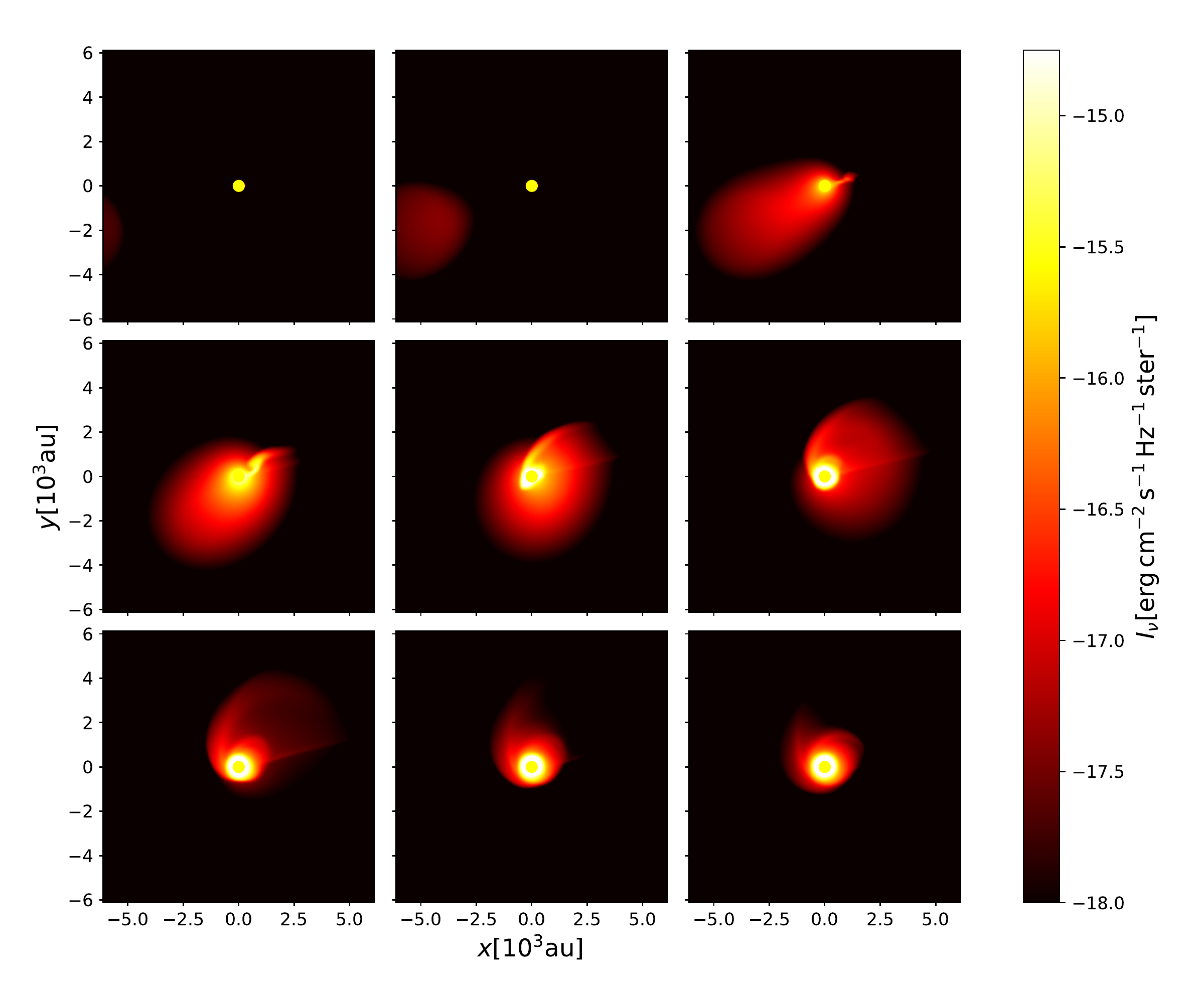}
  }
  \caption{\label{fig-model-I3-scat9}As Fig.~\ref{fig-model-I1-scat9} but now
    for model I3 which has
    $R_{\mathrm{cloud}}=1.2\,b_{\mathrm{crit}}=2662\,\mathrm{au}$ and
    $b=1.0\,b_{\mathrm{crit}}=2218\,\mathrm{au}$ (i.e.~like model A3, but
    now isothermal). \revised{Note the different axis
  size compared to Fig.~\ref{fig-model-I1-scat9}.}}
\end{figure*}

\revised{Finally, in Fig.~\ref{fig-model-I3-scat9} the results of model I3 are
  shown. Compared to models I1 and I2 the spiral and arc features are less
  prominent, which is to be expected, since the ratio
  $R_{\mathrm{cloud}}/b_{\mathrm{cloud}}$ is larger for this model, so that the
  role of the angular momentum of the cloudlet with respect to the star
  is less than in models I1 and I2. As a result, the cloudlet-star
  interaction is more similar to classical Bondi-Hoyle accretion. Indeed,
  one can see a Bondi-Hoyle tail emerging in the top-right panel. This
  tail remains visible in the scattered light images, until the density
  drops too low.}

\revised{In general one can see that for the isothermal models the arc-like
  shape of the nebulosity is less pronounced than for the adiabatic models.  The
  reason is the thermal expansion of the cloudlet. } For the isothermal models
the cloudlet is not pressure-confined. It therefore expands as it approaches the
star. As a result, the arc shaped reflection nebula is only relatively
short-lived, as the cloudlet quite rapidly dissipates. \revised{In contrast, in
  the adiabatic model the cloudlet remains pressure-confined by the warm neutral
  medium. After passing by the star, the tidal stretching creates an arc. In
  the adiabatic case this arc remains geometrically narrow, while in the isothermal
case the arc blows up and dissipates.}

\revised{The thermal expansion of the isothermal models during the flyby} also
means that the dynamic behavior of the cloudlet is much less well described with
ballistic trajectories. The analysis of Section \ref{sec-estimates} is therefore
less applicable to the isothermal models as it is to the adiabatic models.

On the other hand, the isothermal models are more conducive to forming a
disk. This has two reasons. One is that the cloudlet expands as it approaches
the star, and thus more matter may get captured. Secondly, the captured material
cannot get shock-heated, nor can mixing with the Warm Neutral Medium, and thus
cannot build up sufficient pressure to \revised{counteract} the formation of a disk.

\subsection{Discussion of adiabatic versus isothermal results}
\revised{The adiabatic models describe the case of a pressure-confined cloudlet
  passing by the star. They lead to arc-shaped and/or tail-shaped nebulosity
  around the star. But they are less efficient in capturing gas. The isothermal
  models, on the other hand, easily capture material from the cloudlet, and also
  produce Bondi-Hoyle like tails. But they are not efficient in forming
  arcs. Reality probably lies in between these two extreme cases. At large
  distances from the star the clouds may be pressure-confined by the warm
  neutral medium. But whether this is a long-lived state depends on the
  complexities of the phase transition between the cold and warm medium.
  However, once such a confined cloudlet gets tidally disrupted by the star,
  some of this cold cloud material gets compressed and may thereby start to cool
  more efficiently. This cooling will act against the adiabatic heating, and
  thus prevent the ``bounce back'' we see in the adiabatic models. If the
  cooling is fast enough, this material may stay cold during this compression,
  and can thus efficiently form a disk (or merge with an already existing
  disk). At the same time, if the cooling is not too fast, the pressure
  confinement of the tidally stretched flyby material may still keep the arc
  narrow. The exact outcome will depend on the intricacies of the
  heating/cooling physics.}

\revised{If, on the other hand, cloudlets are not pressure confined, and instead
  they simply appear and disappear due to the supersonic turbulence in the
  molecular cloud complex, then the isothermal models may be not a bad
  description of the cloudlet capture process. Some of the material swings by
  the star, while the other part forms a disk. Arcs will then be rarer, but
  sometimes still visible. However, as the material approaches the star, the
  luminosity of the star itself will start to heat the material. The next step
  in the modeling would then be to include this effect. We will study this in a
  follow-up paper (K\"uffmeier, Goicovic \& Dullemond in prep).}

\section{Observations}
\label{sec-obs}
\subsection{AB Aurigae (Transition Disk)}
\begin{figure}
  \centerline{\includegraphics[width=0.45\textwidth]{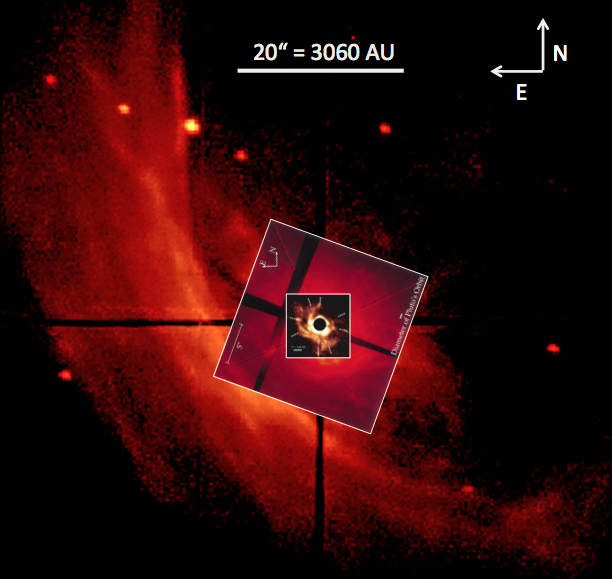}}
  \caption{\label{fig-abaur-scatlight}Composite image of AB Aurigae.  The large
    scale image is from \citet{1999ApJ...523L.151G}, taken with the Uni Hawaii
    2.2 m Telescope at $\lambda=0.647 \mu$m by P.~Kalas.  The medium scale inset
    is from the same paper, taken with the STIS instrument on the Hubble Space
    Telescope at $\lambda=0.57 \mu$m.  Both images were rotated with respect to
    their published form to put north up. The smallest inset is from
    \citet{2004ApJ...605L..53F}, taking with the CIAO instrument on the Subaru
    telescope in the H band ($\lambda\simeq 1.6 \mu$m).  The scale bar is 20''
    which for $d=153$ pc \citep[Gaia DR1 data release: ][]{2016A&A...595A...2G}
    amounts to 3060 AU. The image shows the disk in the center out to at least
    430 AU and a fainter much larger arc-shaped envelope with nearest approach
    to the star on the sky of about 1300 AU, but extending out to at least 6000
    AU. All images used with permission of the authors of the referenced
    papers.}
\end{figure}

A promising candidate for such a `late stage asymmetric cloudlet capture'
scenario is the star AB Aurigae. This star is among the brightest and
nearest Herbig Ae stars with spectral type A0, an estimated mass of
$M_{*}=2.4\,M_{\odot}$ \citep{1998A&A...330..145V}, \citep[but see][for
another estimate of $M_{*}=3.2\,M_{\odot}$]{1992ApJ...397..613H}, an age
estimate of between 2 Myr \citep{1998A&A...330..145V} and 4 Myr
\citep{2003ApJ...590..357D}, and a distance of 153 pc \citep[Gaia DR1 data
release:][]{2016A&A...595A...2G}. 
It was classified as a ``pre-transition
disk'' source by \citet{2010ApJ...718L.199H}, and as member of ``group I''
by \citet{2001A&A...365..476M}. Optical imaging of AB Aurigae revealed not
only the circumstellar disk at scales out to 450 AU, but also a large
arc-shaped reflection nebula to the south-east at scales ranging from
$\sim$1300 AU out to $\sim$6000 AU \citep{1995AJ....109.1181N,
  1999ApJ...523L.151G} (see Fig.~\ref{fig-abaur-scatlight}). The mass in the
arc-shaped nebula was estimated by \citet{1995AJ....109.1181N} to be at
least $2\times 10^{-7}M_{\odot}$, but perhaps more (an upper limit was not given). 
According to \citet{2005ApJ...621..853S}
there is evidence from DCO$^{+}$ measurements with the IRAM 30 m telescope,
as well as IRAS 60 $\mu$m measurements, of cold gas and dust material
extending all the way out to 35,000 AU. 
The archival DSS images also show the reflection nebulosity in a shape of 
a streak extending along the east-west direction up to about 30,000 AU 
\footnote{This streak feature can also be found in 
https://apod.nasa.gov/apod/ap170311.html.}. 
The disk itself, as seen in 
scattered light \citep{1999ApJ...523L.151G, 2004ApJ...605L..53F}, displays
irregular spiral structures at scales of $\sim$ 150$\cdots$450 AU (see
Fig.~\ref{fig-abaur-scatlight}). At smaller scales the disk structure shows
two roughly concentric 
rings in the H band, one with a radius of about 
100 AU and an irregular ringlike/elliptic structure with a radius of about 35$\cdots$60 AU 
\citep{2008ApJ...679.1574O, 2011ApJ...729L..17H}. The outer ring was also
seen at 1.3 mm dust continuum \citep{2005A&A...443..945P,
  2012A&A...547A..84T}, albeit at a slightly larger radius of about 140
AU. These observations show this dust ring to be lopsided to the west, with
an intensity contrast of about 3.

Resolved observations of the $^{12}$CO 2-1 
line by \citet{2012A&A...547A..84T} give an intriguing,
yet confusing picture of the dynamics of the system, which we reproduce in
the text below for convenience. At large scales the $^{12}$CO 2-1 obtained
with the IRAM 30 meter telescope reveal only vague structures. But if these
structures are real, then gas in the huge arc appears to show a radial
velocity change along the arc from $\Delta v\simeq -0.57$ km/s (to the east,
compared to the systemic velocity) to $\Delta v\simeq +0.85$ km/s (to the
south), which is substantial compared to the Kepler velocity at 1400 AU of
$v_{\mathrm{K}}(1400\mathrm{AU})=1.23$ km/s, but consistent with an elliptic
or hyperbolic keplerian orbit. Closer in, at the location of the spirals,
the $^{12}$CO 2-1 data show two velocity components, one of which is also
seen in $^{13}$CO 2-1 data by \citet{2005A&A...443..945P} and is believed to 
be from the disk, the other is 
not seen in the $^{13}$CO 2-1 data and is clearly from the spiral arms. The
disk velocity component is consistent with a disk at an inclination of
$i=23^\circ$ (where $i=0^\circ$ is face-on) and a position angle of the rotation axis of
$\theta=-31.3^\circ$ from north, assuming that the disk rotates
counter-clockwise. However, the spiral component in the $^{12}$CO 2-1 data
appears to behave exactly oppositely: if these data are interpreted as
circular keplerian rotation in the same direction on the sky as the disk
component, one would obtain an inclination of $i=-20^\circ$. Alternatively one
could interpret this as counter-rotating gas at an inclination of
$i=+20^\circ$. 
At even smaller scales (insize of the 140 AU dust ring) 
gaseous spiral arms have been found \revised{by \citet{2017ApJ...840...32T}.}
The origin of these spiral arms is not clear, but they may be induced by an unseen companion or planet.  

Interpreting these data is not straightfoward, but there is little doubt
that the disk of AB Aurigae is at this moment being fed with fresh material
from its surroundings, as has been reported numerous times in the
literature. 
The arc is, in our opinion, a stream of gas+dust stretched along its
fly-by trajectory as a result of the tidal forces of AB Aurigae's
gravitational field. The free-fall time at the closest approach of 1300 AU
is 3400 years. With a conservative estimate of the velocity along this
trajectory of 1 km/s, the gas in the stream flows along its entire observable
path in about 20,000 years, which is less than 1\% of the age of the
star. The formation time scale of a star of this mass is of the order of
$10^5$ years, meaning that any material that is directly related to the
original cloud core must have already accreted or escaped the
system long ago, by a factor of 20 or more in time. The material that is
currently observed falling onto the disk (or flying by) is therefore
unrelated to the original star formation event. It must be from a random cloud
fragment of the larger scale molecular cloud complex. 
The extended CO as well as a reflection nebulosity in $>$10,000 AU scale 
support this picture.

\subsection{HD 100546 (Transition Disk)}
\begin{figure}
  \centerline{\includegraphics[width=0.45\textwidth]{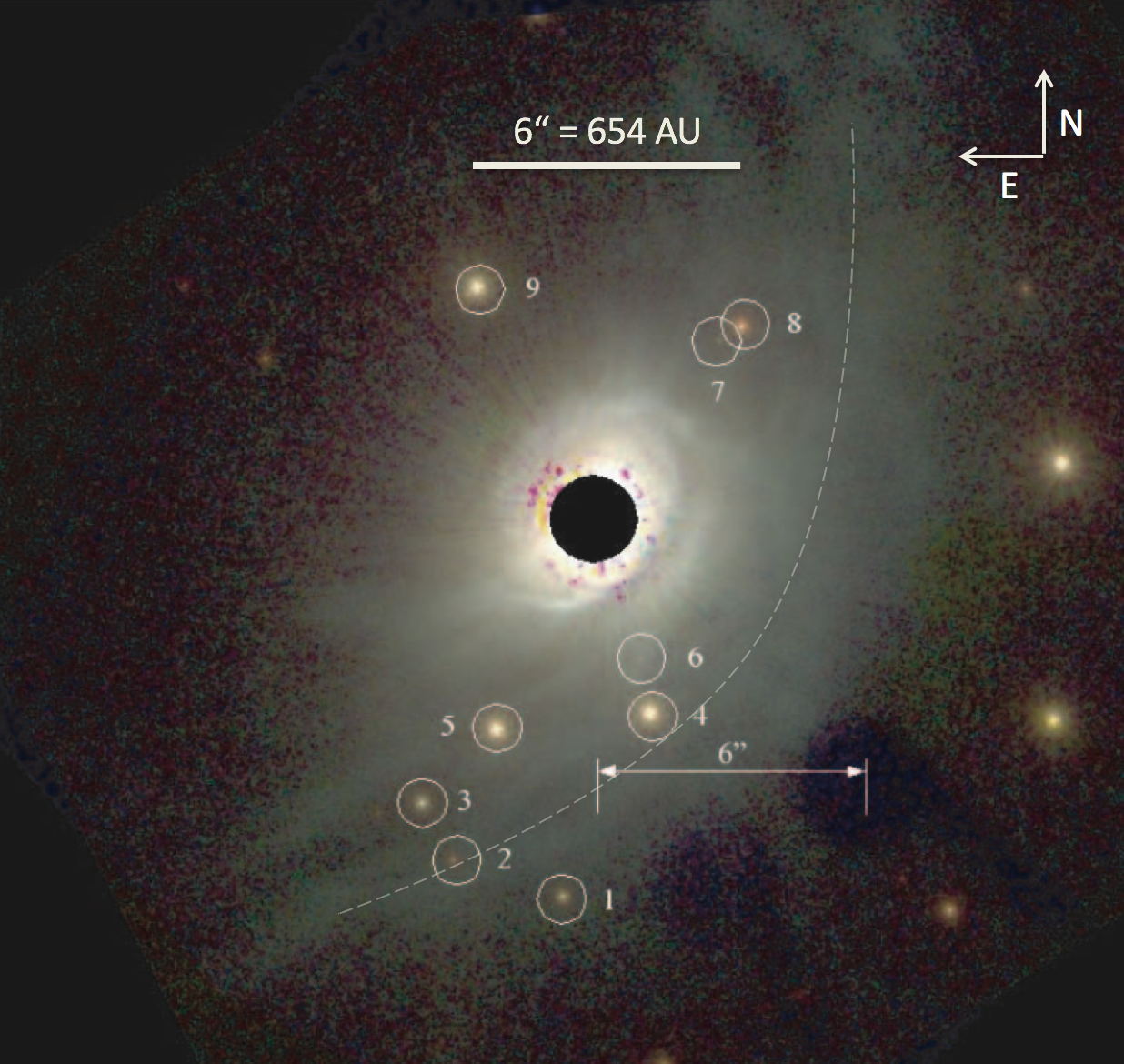}}
  \caption{\label{fig-hd100546-scatlight}Image of HD100546 from
    \citet{2007ApJ...665..512A} (their Figure 5) with RGB color coding
    according to R=B-band, G=V-band and B=I-band. The scale bar is 6'' which
    for $d=109$ pc \citep[Gaia DR1 data release: ][]{2016A&A...595A...2G})
    amounts to 654 AU. The image shows the disk in the center out to at
    least 300 AU, and a much fainter, much larger arc-shaped envelope out to
    1000 AU going from south, via south-west to north-west (a dashed line
    was added to the image to indicate its position, as it is rather faint).
    Image used with permission of the authors of the referenced
    paper.}
\end{figure}

Hubble Space Telescope (HST) observations by \citet{2001AJ....122.3396G,
  2007ApJ...665..512A} revealed that the Herbig Ae star HD 100546 is
surrounded by a large disk with a radius of about 300 AU, as well as by
tenuous envelope material out to about 1000 AU, in spite of its old age of
10 Myr \citep{1998A&A...330..145V}. The disk has $m=2$ spiral structure globally \citep{2001AJ....122.3396G, 2007ApJ...665..512A}, albeit not nearly as strong 
as in sources such HD 100453 \citep{2017A&A...597A..42B}. 
\citet{2006ApJ...640.1078Q} suggest that the
spiral patterns can be explained by illumination from the star if the disk
is warped. 
HD 100546 has also been known to feature an inner hole of about 13 AU
radius \citep{2003A&A...401..577B, 2005ApJ...620..470G}, later confirmed in 
the direct imaging at optical and infrared wavelengths 
\citep{2016A&A...588A...8G, 2017AJ....153..264F}, thus clearly making 
it a Transition Disk.
It must also have at least some material close to the
star (around 0.5 AU distance, where the dust sublimation radius is), owing
to its observed near-infrared excess. Compared to AB Aurigae, however, the
near-infrared excess of HD 100546 is much weaker. This suggests that the
inner disk is substantially less dense in this source. Like in the case of
AB Aurigae, the larger scale envelope material appears to be arranged in an
arc-like shape around the south-west part of the disk at scales of up to 1000 AU, most clearly seen
in Fig.~5 of the paper by \citet{2007ApJ...665..512A}, reproduced in
Fig.~\ref{fig-hd100546-scatlight}.

Assuming that the spirals are trailing and are part of the disk itself, the
disk rotates counter-clockwise on the sky. The fact that the CO 3-2 first
moment map of \citep{2014ApJ...791L...6W} shows a clear rotation profile
with blueshifted emission on the south-east and redshifted emission on the
north-west, means that the inclination of the rotation axis of the disk is
pointing to the north-east, meaning that the near side of the disk is on the
south-west. More precise analysis \citep{2014ApJ...791L...6W} gives an
inclination of $i=44$ from pole-on, with the rotation axis of the disk
pointing at a position angle on the sky of 56$^\circ$ counterclockwise from
north. In the scattered-light image of \citet{2007ApJ...665..512A} it
appears that there is a dark lane on the near-side, which is consistent with
the shape expected from an inclined disk with a bright scattered light
surface on both the front and the back sides of the disk. If this
interpretation is true, then this confirms the inclination and position
angle inferred from the CO 3-2 data. 
The detailed look at the CO 3-2 first moment map provides a hint of 
a systematic deviation from the Keplerian velocity at the location of 
the south-west spiral at $\sim$350 AU, though only by one velocity-resolution 
element ($\Delta v\simeq -0.21$ km/s). Inside 100 AU, non-Keplerian gas 
motion is suggested, which can be accounted for by the inner disk misaligned 
to the outer disk, or a radial flow of gas \citep{2017A&A...607A.114W}. 

The system therefore shows a similarity to AB~Aurigae in its circumstellar structure, and looks like a more evolved counterpart of AB~Aurigae. 
In particular, such a longevity of the large-scale envelope implies the star-forming environment rich in cloud fragments and a higher possibility of mass supply from the envelope onto the disk at later epochs than the original, main accretion phase of the protostar.

\subsection{Other Transition Disk sources}
There are other Transition Disk sources with reflection nebulosity, though most
are faint and do not show arc-like structures. In the case of HD 142527, a faint arc can perhaps be seen (see archival DSS images\footnote{http://archive.eso.org/dss/dss}) and a 500 AU-scale arm is found in CO observations \citep{2014ApJ...785L..12C}, 
but it is certainly not massive enough to have a
direct causal relationship with the tilted outer disk of HD 142527 \citep{2015ApJ...798L..44M}. If the
tilted disk around this star was caused by a cloudlet capture event, it must have
taken place long enough ago, that no traces of this event are left. The star HD
97048 has quite some reflection nebulosity around it, and filamentary, spiral-like structures have been found at $\sim$500~AU  \citep{2007AJ....133.2122D}, but also here no clear
arcs can be identified.
RY~Tau illuminates the surrounding nebula \citep{1995AJ....109.1181N}, and the presence of an envelope has also been confirmed in the high resolution imaging of the inner 100 AU \citep{2013ApJ...772..145T}, but again the shape of the nebula cannot be recognized as an arc.
Other transitional disks, for instance those listed in Table 1 of \citet{2014prpl.conf..497E} show no nebulosity based on the search using the DSS archival images, suggesting that the frequency of $>$1000 AU-scale nebulosity is $\sim$10\%. 
Note that the tenuous envelope of HD~100546 detected with HST cannot be identified in the DSS, and high-sensitivity observations for a statistically-meaningful sample are required to conclude the occurrence rate of such nebulosity.
The probability of having a long-lived nebulosity seems to be lower for lower-mass stars than $\sim$$2\,M_{\odot}$ since no T Tauri stars show reflection nebulae in the DSS search. 
However, for GM~Aur \citep[$0.84\,M_{\odot}$][]{2000ApJ...545.1034S}, the HST observations \citep{2003AJ....125.1467S} revealed the kinked, ribbon-like feature stretching toward northeast out to about 1700 AU from the star.

\subsection{FU Orionis sources}
FU Orionis stars are normally associated with reflection nebulosity unless they suffer from the large extinction by the parent molecular clouds. 
This is quite reasonable since most FU Orionis sources have prominent envelopes because of their youth, but it would be worth pointing out that at least two of them, FU~Ori and Z~CMa, are surrounded by the nebulae of 1000--10000 AU-scale in the shape of arcs \citep{1995AJ....109.1181N}.
When looking into their inner ($<$1000 AU) regions, the spatial structures are strikingly inhomogenious, which have been observed in scattered light in near-infrared \citep{2016SciA....200875L}. 
% H or Ks bands
FU Ori shows an arm structure in the eastern side at 50--500 AU from the star, without a counter arm in the west. 
Z~CMa also has an arm-like structure $\sim$300 AU south of the star, in addition to a stream extending toward southwest. 
The companion stars are known for these stars, which complicates the interpretation of circumstellar structures, but in any case, such non-axisymmetry can be linked to the episodic accretion events of this class of objects. 
The other two FU Orionis sources in \citet{2016SciA....200875L} also show arms although the larger-scale envelopes do not look like arcs.
Significant arc structures were found toward three stars out of 22 including "FUor-like" listed in \citet{2014prpl.conf..387A} in the DSS images: FU Ori,
Z CMa and V646 Pup.

\section{Discussion}
\label{sec-discussion}

\subsection{Transitional disks caused by cloudlet capture?}
Here we speculate whether cloudlet capture may be related to the class of large
Transitional Disks, and might explain why some of them appear to be extremely
warped. Transitional disks (TDs) are protoplanetary disks with a large
cavity. The gas and dust in this cavity has been removed, or at least strongly
suppressed with respect to the outer disk. It is still a matter of debate which
process has carved out this cavity. Many TDs still have a small inner disk
inside the cavity \citep[e.g.][]{2007ApJ...664L.107B, 2010ApJ...717..441E}, and
are sometimes called ``pre-Transition Disks''. This leads to an ``inner disk -
gap - outer disk'' geometry. The gap is, however, much more radially extended
than one would expect for a gap carved out by a planet. A multi-planet system
might open up a gap of that size, but it is hard to explain the depth of the gap
with such a scenario \citep{2011ApJ...729...47Z}. A binary stellar companion may
play a role. For instance, the star HD 142527, which features a prominent TD,
has been resolved to consist of the main star and its M-dwarf companion at about
10 au projected distance \citep{2012ApJ...753L..38B}, i.e.\ well within the 120
au gap of the disk. The inner disk appears to be resolved in recent images with
SPHERE \citep{2017AJ....154...33A}, indicating a radius of a few au.

Recently it was found that in some TDs the inner disk appears to be strongly
tilted with respect to the outer disk \citep{2015ApJ...798L..44M}.  The striking
two dark spots seen in scattered-light images of the outer disk of HD 142527 are
naturally explained as being caused by the shadow cast by a heavily inclined
($\sim 70^\circ$) inner disk. The same phenomenon was observed in the disk of HD
100453 and also here the shadow cast by a heavily inclined ($\sim 72^\circ$)
inner disk matches the observations \citep{2017A&A...597A..42B}. This
misalignment of the angular momentum vectors of the inner and outer disks in
these two objects is so extreme (nearly perpendicular), that it is hard to
imagine this to be caused by disk-interal processes or planetary objects.

In the case of HD 142527 the close-in binary companion may be orbiting out of
the plane of the outer disk, which would explain why the inner disk is so
strongly inclined. If the outer disk of this system is, however, of primordial
nature, then this raises the question: why would the binary companion (which
would be formed, one would think, from the same primordial disk) move on such
a wildly out-of-plane orbit?

In the case of HD 100453 there appears to be a companion M star at 120 au
projected distance, i.e.~outside of the main star+disk system. Also this
companion will likely have a strong dynamical influence on the disk, and may be
responsible for the strong $m=2$ spiral feature seen in this disk
\citep{2016ApJ...816L..12D}. If this companion orbits out of the plane of the
disk, it might have caused the outer disk to precess and cause the misalignment
of the inner and outer disk. But here again it is puzzling why the binary and
the disk are so inclined with respect to each other.

One possible explanation was suggested by \citet{2017MNRAS.469.2834O}, who show
that a binary companion inside the gap may lead the disk to undergo a secular
precession resonance. This would make an initially in-plane binary+disk
configuration to become mutually inclined.

Here we propose an alternative explanation: that these stars or binaries have,
at some point {\em after} their formation, captured a low mass cloudlet from
what is left of the surrounding clumpy giant molecular cloud (GMC) in which they
were born. This cloudlet capture process replenishes the mass in the disk (or wraps
an entirely new second generation disk around the system), but typically with an
angular momentum axis different from that of the primordial disk. This is
because the cloudlet originates from a different part of the GMC, and is thus
dynamically unrelated to the original collapsing cloud core that produced the
star and its primordial disk. 

If at the time of the cloudlet capture event a small primordial disk still existed,
the formation of the secondary disk would lead to the inclined inner/outer disk
geometry. If the primordial disk is large, the captured cloudlet material would
violently hydrodynamically interact with the primordial disk, possibly tilting
its angular momentum axis. If the star is in fact a binary system (such as in HD
142527), then the circumbinary disk would likely be misaligned with the binary
orbital plane. Any gas accretion from the circumbinary disk into the inner
system would then likely get ``reoriented'' to the binary plane, again yielding
an inclined inner/outer disk geometry.

A main caveat of this scenario is the required mass of the cloudlet that is
captured, see Subsection \ref{sec-caveat-cloudmass}. Also, in a process as
complex as star formation in clusters it may not be possible to cleanly
distinguish between what constitutes primordial infall and what is a late-stage
cloudlet capture event, as we will discuss in Section
\ref{sec-caveat-primordial-collapse}.

\subsection{Can cloudlet capture events trigger FU Orionis outbursts?}
FU Orionis outbursts are relatively sudden, long-lasting outbursts of accretion
activity in protostellar disk sources \citep{1977ApJ...217..693H,
  1991ApJ...383..664K}. Many of these objects are heavily embedded in molecular
clouds while some are more revealed objects. The outbursts are usually believed
to be the result of an instability that periodically drives the disk from a
long-lasting cold ``low'' state to a (relatively) short duration hot ``high''
state, and back again \citep[e.g.][]{1991ApJ...383..664K, 2001MNRAS.324..705A,
  2009ApJ...694.1045Z, 2010ApJ...713.1143Z}. In the low state the disk cannot
transport as much matter as it is being fed from outside (either from the outer
disk of from an infalling envelope), and therefore starts to sequester this
mass. Once the surface density of the disk reaches a critical value, the disk
switches to the high state, rapidly flushes the sequestered mass onto the star,
and returns to the low state. The low/high state are related to temperature
through the ionization degree in the disk, as a lack of free ions and electrons
leads to the formation of ``dead zones'' which are inefficient in driving
accretion.  The gravitational instability might play a role in triggering the
transition from low to high. \citet{2007MNRAS.381.1009V, 2010ApJ...719.1896V}
show that the interplay between continued feeding from the envelope causes the
disk to display bursts of accretion driven by the gravitational instability,
even if the turbulent viscosity of the disk is so low as to be unimportant.

The reflection nebulosity around several FU Orionis star systems
\citep{2016SciA....200875L} suggests that indeed FU Orionis outbursts could
be related to an outside feeding of the disk. While this might be still
part of the original star formation cloud collapse, these tail-like and
arc-like features are suggestive of cloudlet capture events occurring.
If this is the case, some FU Orionis objects may be ``rejuvenated'' disks.

We are, however, aware that this is very speculative. In particular, it
would presumably require a relatively massive cloudlet capture event to take
place. Such a cloudlet may even be optically thick and not show the kind of
reflection nebula shapes discussed in this paper. It would be interesting,
however, to further explore direct observational evidence of a link between
infall onto the disk and the occurrance of FU Orionis outbursts.

EX Lupi (``EXOr'') outbursts are similar to FU Orionis outbursts, but
they last much shorter. The star EX Lupi is the archetype of this class
of objects and had its most recent outburst in 2008 lasting about half
a year. These outburst are repetitive on time scales of tens of years.
Given that this repetition rate is on a much shorter time scale than
cloudlet capture events, it is unlikely that each EXOr outburst is associated
with such an event. But like with FU Orionis outbursts, a single cloudlet
capture event might lead to ``overloading'' of the disk, which leads to
multiple outbursts on short time scales \citep{2012MNRAS.420..416D}.

\subsection{Disk replenishment via cloudlet capture: the issue of cloudlet mass}
\label{sec-caveat-cloudmass}%
In order for a late-type cloudlet capture event to significantly replenish
the protoplanetary disk of a star, and thereby possibly tilt it to another
rotation axis and/or trigger a FU Orionis activity, the mass of the cloudlet
must be sufficiently large. Since often part of the cloudlet flies by and
only a fraction of the cloudlet gets accreted, this mass would have to be
accordingly larger to compensate for this inefficiency of capture.

In our models we rely on the mass-radius relation of cloudlets as given by
Eq.~(\ref{eq-mass-radius-relation}). This relation from
\citet{2010A&A...520A..17K} was based on data from \citet{2004Ap&SS.292...89F}
which go all the way down to scales of $5\times 10^{-3}\,\mathrm{pc}$ and cloudlet
masses of $10^{-4}\,M_{\odot}$. This relation has a scatter of about a factor of
$10^2$ in cloudlet mass at scales where the most data is available.

Arc-shaped features are expected to be most prominent for cloudlets that have radii
similar to $b_{\mathrm{crit}}$. For a star of $2.5\,M_{\odot}$ and a relative
velocity of $v_{\infty}=1 \mathrm{km/s}$ this would be about 2200 au, which,
according to the mass-radius relation, amounts to a mass of
$M_{\mathrm{cloud}}\simeq 1.5\times 10^{-3}\,M_{\odot}$. The optical depth of
the cloudlet at $\lambda=0.65\,\mu$m would be about $0.1$.

This cloudlet mass is, however, on the low side for replenishing and old (or
creating a new) disk around the star. For instance, the outer disk of the
Transition Disk star HD 142527 was estimated by \citet{2012ApJ...754L..31C} to
have a mass of about 0.1 $M_{\odot}$. For AB Aurigae the disk mass estimated
from dust continuum is between $0.01\cdots 0.04\,M_{\odot}$
\citep{2012A&A...547A..84T}. For HD 100546 this is between
$1\cdots 5\times 10^{-3}\,M_{\odot}$
\citep{2003A&A...398..607D, 2011A&A...530L...2T}, although
if we scale the estimated dust mass up with an assumed gas-to-dust ratio of 100
one would obtain a gas mass of $5\times 10^{-2}\,M_{\odot}$
\citep{2010A&A...511A..75B, 2011A&A...530L...2T}.

Assuming instead a cloudlet mass of 0.1 $M_{\odot}$, the mass-radius relation would
yield a cloudlet radius of about 6 $b_{\mathrm{crit}}$ and an optical depth of
$0.15$.  Such a large cloudlet would presumably not yield clearly identifiable
arc-shaped reflection nebulosity, although it might generate Bondi-Hoyle
tail-shaped features.

The mass-radius relation has, however, a spread in mass. If we would take an
0.1 $M_{\odot}$ cloudlet of radius $b_{\mathrm{crit}}$, a clear arc will be formed, but
it will likely be optically thick. The cloudlet capture event would then look
more like a class I object instead of a class II object.

\subsection{Cloudlet capture versus the primordial collapse}
\label{sec-caveat-primordial-collapse}%
In recent years numerous examples of misalignment in very young stellar objects
have been found. For example, \citet{2016ApJ...820L...2L} find misalignments
between the outflow axes in binaries in the Perseus molecular cloud.
\citet{2016ApJ...830L..16B} find a circumbinary disk around the class I binary
star Oph IRS 43 that is misaligned with the orbit of the binary. Such findings
show that star formation is not simple and that the angular momentum axis of
infalling material varies with time, even already during the main cloud collapse
phase. Numerical star formation modeling seems to confirm this
\citep{2010MNRAS.401.1505B,2018MNRAS.475.5618B}. It is therefore very well possible that the
misalignment of the outer/inner disks of several Transition Disks could be
directly inherited from the very early phases of the formation of the system.
\revised{In the models of \citet{2018MNRAS.475.5618B} this is most strikingly seen
  in his Fig.~2, where a binary star is formed with an outer disk misaligned
by 75 degrees with with the inner circumbinary disk.}
In that case, no late-stage cloudlet capture is necessary to create the strong
misalignment, and the outer disk is then primordial, not secondary. \revised{But of
  course,} a
clear distinction between ``primordial misalignment'' and ``misalignment due to
late-stage cloudlet capture'' may be hard to define since even the primordial
accretion phase may be messy. Full zoom-in star formation simulations of the
kind of \citet{2017ApJ...846....7K} over a few million years may give the answer
whether late stage cloudlet capture events can affect the protoplanetary disk's
axis. Such simulations will also show whether such secondary accretion events
would produce larger disks than the primary events, because of large angular
momentum resulting from the asymmetric approach of the cloudlet toward the
star. This could be a criterion to distinguish the two.

\subsection{Cloudlet capture in other contexts}
The phenomenon of ``cloudlet capture'' also plays a role in the context of
supermassive black holes in the center of galaxies, including our own.
\citet{2006MNRAS.373L..90K} suggest that supermassive black holes may feed
themselves through random infalling cloudlets with near-zero angular momentum. A
somewhat larger angular momentum (i.e.\ impact parameter) would lead to the
formation of an eccentric disk around the black hole and possibly the formation
of stars in this disk \citep[e.g.][]{2008Sci...321.1060B,2016MNRAS.455.1989G}.
In case of a binary black hole system the randomly oriented circumbinary disk
that is formed in such an event \citep{2014MNRAS.445.2285D} might play a role
in reducing the orbital separation of the black holes and thus resolving the
``last parsec problem'' of supermassive black hole merging.

While qualitatively the two scenarios are similar, the main difference lies in
the ratio of the gas temperature to the virial temperature. At the scale of
$r=0.1\,b_{\mathrm{crit}}$ around a T Tauri or Herbig Ae star the virial
temperature $T_{\mathrm{vir}}=\mu m_pGM_{*}/k_Br$ is of the order of 3000 K. If
the gas has a temperature of 30 K, for instance, this leads to a disk vertical
scale height of $h_p/r=c_s/\Omega_Kr=\simeq 0.1$ (where $c_s$ is the isothermal
sound speed and $\Omega_K$ the Kepler frequency). For molecular disks around
supermassive black holes the ratio of the gas temperature to the virial
temperature is orders of magnitude smaller due to the enormous depth of the
gravitational potential well. The disks that are formed are therefore also
geometrically much thinner, with values of the order of $h_p/r\simeq
10^{-3}\cdots 10^{-2}$. The difference between the two cases is qualitatively
the strongest during the approach of the cloudlet to the star / black hole: the gas
temperature plays a much larger role in the case of a cloudlet approaching the
star, while in the case of clouds accreting onto a supermassive black hole the
approach is more ballistic.

\subsection{Scalability of the model}
Because in our hydrodynamic model we have used simple equations of state (only
adiabatic and isothermal, no cooling), the results of the calculations can be
relatively easily scaled to other stellar masses, luminosities and cloudlet masses
and sizes. Also the mass-size relation (Eq.~\ref{eq-mass-radius-relation}) does
hot have to be strictly adhered to. The only non-scalable (dimensionless)
parameters are $b/b_{\mathrm{crit}}$, $R_{\mathrm{cloud}}/b_{\mathrm{crit}}$ and
the Mach number ${\cal M}\equiv v_{\infty}/\sqrt{k_BT_{\mathrm{cloud}}/\mu
  m_p}$.  And of course the initial position of the cloudlet relatively close to
the star for the isothermal models, in units of $b_{\mathrm{crit}}$. The Mach
number is likely to play only a limited role in the adiabatic models, given that
the cloudlet is pressure confined by the warm neutral medium.

\subsection{Caveats and future work}
The simple modeling setup with a fixed grid that is used in this paper is not
suited for studying the actual formation of a secondary disk and/or the
cloud-disk interaction. Adaptive mesh refinement is unavoidable to get
sufficient spatial resolution to resolve the vertical scale of the
protoplanetary disk. \revised{Also a more realistic equation of state will be
  required to make quantitative conclusions. In a follow-up paper (K\"uffmeier,
  Goicovic \& Dullemond in prep) we will present the results of cloudlet capture
  models simulated with the AREPO moving mesh code, in which we study,
  among other things, the properties of the resulting secondary disks, and the
  effects of more realistic equations of state. Further down the road it will be
  necessary to replace the simple initial conditions with a realistic
  environment: Starting with a large scale star formation model, and
  performing zoom-in simulations on individual stars, and modeling these well
  beyond their initial collapse phase.}

\revised{Magnetic field may play a critical role as well. It is likely that low
  mass cloudlets are more magnetized (i.e.\ have a higher magnetic over gas
  pressure ratio) than the much more massive star-forming cloud cores. It is
  unclear how this may change the results of cloudlet capture. Magnetic pressure
  may give the cloudlet an ``adiabatic-like'' behavior, which would suppress
  capture of the gas, while magnetic tension may, on the contrary, extract
  angular momentum from the captured material, enhancing the capture of gas.
  And whether arc-shaped or tail-shaped features will be formed by the
  non-captured material, when magnetic fields are involved, is entirely
  unclear. On the other hand, it is known that filamentary structures
  spontaneously form when magnetic fields start to dominate over gas
  pressure. An extreme example of this are the coronal loops in the solar
  chromosphere and corona. But similar effects also occur in molecular cloud
  complexes \citep{2013A&A...556A.153H}.  Clearly, MHD modeling will be needed
  to answer these questions.}

\section{Conclusions}
Cloudlet capture is a process that can replenish the circumstellar environment of
young stars even at a relatively late time after formation, as long as the star
is still travelling within the star formation region and the material of the
giant molecular cloud complex has not yet fully dissipated.  As opposed to the
``classical'' Bondi-Hoyle-Lyttleton accretion, cloudlet capture involves cloudlets
or cloud filaments that are of similar size as, or smaller than, the critical
impact parameter $b_{\mathrm{crit}} = GM_{*}/v_{\infty}^2$, with $v_{\infty}$
being the approach velocity. In such a case, the cloudlet brings a lot of angular
momentum into the process. If the cloudlet passes by close enough to the star (with
impact parameter similar to, or smaller than $b_{\mathrm{crit}}$) this can lead
to the capture of part of this cloudlet followed by the formation of a secondary
disk or the replenishment of an already existing disk. The other part of the
cloudlet will pass by in an arc. Some of this material may, in fact, still accrete
if parts of the arc are elliptic instead of hyperbolic. Searching for such
signatures around young stars may teach us about the frequency of such events,
which gives constraints on the process of star formation and may help explain
some of the more exotic protoplanetary disk sources such as tilted Transition
Disks and FU Orionis outbursting stars.

\begin{acknowledgements}
  We would like to thank Leonardo Testi, Hauyu Baobab Liu, Jonathan Williams,
  Peter Abraham, Agnes Kospal, Klaus Pontoppidan, Giovanni Rosotti, Thomas
  Haworth, Colin McNally, Willy Kley, Antonella Natta, Nienke van der Marel,
  Ewine van Dishoeck, Ralf Klessen, Philipp Girichidis,
  Patrick Hennebelle, Mario van den Ancker, Monika Petr-Gotzens, Adriana Pohl,
  Carsten Dominik, for useful discussions, and Daniel Thun for help with the
  Binac Cluster. The authors acknowledge support by the High Performance and
  Cloud Computing Group at the Zentrum f\"ur Datenverarbeitung of the University
  of T\"ubingen, the state of Baden-W\"urttemberg through bwHPC and the German
  Research Foundation (DFG) through grant no INST 37/935-1 FUGG. This research
  was supported by the Munich Institute for Astro- and Particle Physics (MIAPP)
  of the DFG cluster of excellence ``Origin and Structure of the Universe'', in
  particular the MIAPP workshop ``Protoplanetary Disks and Planet Formation'' in
  June 2017 where many of the above discussions took place which considerably
  helped to improve the argumentation. Part of this work was also funded by the
  DFG Forschergruppe FOR 2634 ``Planet Formation Witnesses and Probes:
  Transition Disks'' under grant DU 414/23-1.  F.G.\ acknowledges support from
  DFG Schwerpunktprogramm SPP 1992 ``Diversity of Extrasolar Planets'' grant DU
  414/21-1. The research of M.K.\ is supported by a research grant of the
  Independent Research Foundation Denmark (IRFD) (international postdoctoral
  fellow, project number: 8028-00025B).
  The hydrodynamics models were done with the PLUTO code
  (\url{http://plutocode.ph.unito.it}). Some supporting models (not shown in
  this paper) were performed with the AREPO code
  (\url{https://wwwmpa.mpa-garching.mpg.de/~volker/arepo/}). Finally, we would
  like to thank the anonymous referee for his/her patience and useful suggestions
  to improve the paper.
\end{acknowledgements}

\begingroup
\bibliographystyle{aa}
\bibliography{ms}
\endgroup

\end{document}